\documentclass[aps,twocolumn,notitlepage,letterpaper]{revtex4-2}
\usepackage[utf8]{inputenc}
\usepackage{graphics,amssymb,amsmath,epsfig,color}
\usepackage{graphicx}
\usepackage{float}
\usepackage{braket}
\usepackage{mathptmx}
\usepackage{times}
\usepackage{mathtools}
\usepackage{bm}
\usepackage{ifpdf}

\newcommand{\bk}{{\mathbf k}}
\newcommand{\br}{{\mathbf r}}

\newcommand{\bp}{{\mathbf p}}
\newcommand{\by}{{\mathbf y}}
\newcommand{\bz}{{\mathbf z}}
\newcommand{\bA}{{\mathbf A}}
\newcommand{\bB}{{\mathbf B}}
\newcommand{\bE}{{\mathbf E}}

\newcommand{\bP}{{\mathbf P}}
\newcommand{\bM}{{\mathbf M}}
\newcommand{\bs}{{\mathbf \sigma}}

\usepackage{amsfonts}

\usepackage{amsmath}
\usepackage{amssymb}
\usepackage{bm}
\usepackage{mathrsfs}
\usepackage{tensor}
\usepackage{dsfont}
\usepackage{graphicx}
\usepackage[caption=false]{subfig}

\usepackage[usenames,dvipsnames]{xcolor}
\usepackage[hyperindex,pdftex, breaklinks,colorlinks = true,linkcolor = blue,urlcolor=blue,citecolor=blue]{hyperref}

\usepackage{physics}

\begin{document}
\title{Bulk versus surface: Nonuniversal partitioning of the topological magnetoelectric effect}

\author{Chao Lei}
\email{leichao.ph@gmail.com}
\author{Perry T.~Mahon}
\email{perry.mahon@austin.utexas.edu}

\author{A.~H.~MacDonald}
\affiliation{Department of Physics, University of Texas at Austin, Austin, Texas 78712, USA}

\thanks{C.~L.~and P.~T.~M.~contributed equally to this work.}

\begin{abstract}
The electronic ground state of a three-dimensional (3D) band insulator with time-reversal ($\Theta$) symmetry 
or time-reversal times a discrete translation ($\Theta T_{1/2}$) symmetry is 
classified by a $\mathbb{Z}_{2}$-valued topological invariant and characterized by quantized magnetoelectric response.
Here we demonstrate by explicit calculation in model $\mathbb{Z}_{2}$ topological insulator thin-films
that whereas the magnetoelectric response is localized at the surface 
in the $\Theta$ symmetry (nonmagnetic) case, it is nonuniversally partitioned between
surface and interior contributions in the $\Theta T_{1/2}$ (antiferromagnetic) case, while remaining quantized.
Within our model the magnetic field induced polarization arises
entirely from an anomalous ${\cal N}=0$ Landau level subspace within which 
the projected Hamiltonian is a generalized Su-Schrieffer-Heeger model whose topological 
properties are consistent with those of the starting 3D model.
We identify a new connection between the ground-state geometry of that 3D model and surface-interior partitioning in thin films.
\end{abstract}

\date{\today}

\maketitle

\section{Introduction}
The realization that material properties can depend qualitatively \cite{Hasan_review_2010,Qi_review_2011} 
on topological data encoded in the crystal-momentum dependence of 
its occupied electronic eigenspace projectors has been an important achievement in modern condensed matter physics.  
The physical import of this mathematical structure was first recognized in 
theoretical studies \cite{TKNN,Avron_1983} 
of the integer quantum Hall effect (IQHE) in which the Hall 
conductivity of two-dimensional band insulators was identified as 
the (quantized) Chern number of the vector bundle of occupied (electronic) Bloch states 
over the Brillouin zone (BZ) torus \cite{avron1984quantum}.
Subsequently other, more subtle, topological invariants of band insulating ground states have been found to be physically relevant. 
Among these, three-dimensional (3D) band insulators 
with time-reversal symmetry (TRS, denoted as $\Theta$) or generalized TRS ($\Theta T_{1/2}$), 
which have a quantized magnetoelectric response \cite{TME_TI_2008,Vanderbilt2009,souza2011chern},
\begin{equation}\label{alpha}
    \alpha^{il}=\left.\frac{\partial P^i}{\partial B^l}\right|_{\substack{\boldsymbol{E}=\boldsymbol{0} \\ \boldsymbol{B}=\boldsymbol{0}}}=\left.\frac{\partial M^l}{\partial E^i}\right|_{\substack{\boldsymbol{E}=\boldsymbol{0} \\ \boldsymbol{B}=\boldsymbol{0}}} =\delta^{il} \alpha,
\end{equation}
have been a prominent focus.  In Eq.~(\ref{alpha}) superscript indices identify Cartesian components,
$\bP$ and $\bM$ are the (electronic) polarization and orbital magnetization, respectively, and $\bE$ and $\bB$ are uniform dc electric and magnetic fields, respectively. 
In both nonmagnetic ($\Theta$ symmetric) and antiferromagnetic (AFM) ($\Theta T_{1/2}$ symmetric \cite{Mong2010}) cases, 
$\alpha=0$ or $e^{2}/2hc$ 
(mod $e^{2}/hc$) and is determined, in part, by the $\mathbb{Z}_2$ classification of the ground state \cite{TME_TI_2008,Vanderbilt2009,souza2011chern}. 
Indeed, an explicit expression has been proposed \cite{TME_TI_2008} that  
relates $\alpha$ to the bulk BZ integral of the Chern-Simons 3-form in the space of occupied Bloch states.
The value of $\alpha$ implied by this bulk expression ($\alpha_{\text{CS}}$), quoted explicitly below, is determined modulo $e^{2}/hc$ by the $\mathbb{Z}_{2}$ index of the vector bundle of occupied Bloch states \cite{TME_TI_2008,Mong2010,monaco2015symmetry,kaufmann2016notes}.

The relationship between $\alpha_{\text{CS}}$ and the physical magnetoelectric response ($\alpha_{\text{me}}$) of finite volume 
$\mathbb{Z}_{2}$ topological insulators (TIs) with surfaces 
is nontrivial \cite{souza2011chern,Nomura2011,Nagaosa2015,vanderbilt_book,mahon2023reconciling}.
In nonmagnetic thin films \footnote{We exclude systems with an integer quantum Hall effect, and distinguish surface and bulk of thin films. We therefore restrict our attention to films with even layer number greater than or equal to four.} with surface magnetic dopants
that have nonzero surface-normal magnetization components that are opposite
on top and bottom surfaces, $\alpha_{\text{me}}$ approaches $\alpha_{\text{CS}}$ in the thick-film limit.
Here we demonstrate by explicit calculation that whereas the quantized magnetoelectric response is localized at the surface 
in the nonmagnetic case, it is nonuniversally partitioned between
surface and interior contributions in the AFM case. 
In both cases the response of the (electronic) charge density to magnetic field is local in the sense that it occurs only in the spatial vicinity of static magnetic moments that break $\Theta$ symmetry. The difference is that these are present throughout the 
film interior in the AFM case.
The dipole moment of the charge density that is induced by the magnetic field in the interior region of an AFM film is nonuniversal, but 
for sufficiently thick films we find that the sum of interior and surface contributions to the dipole moment is quantized (mod $e^2/hc$) 
at either $\alpha_{\text{me}}=0$ or $e^2/2hc$ depending on the $\mathbb{Z}_2$ classification of the bulk ground state; 
in AFM (nonmagnetic) thin films (with magnetic surface dopants) $\Theta T_{1/2}$ ($\Theta$) symmetry is broken and therefore, e.g., $\alpha_{\text{me}}=0$ cannot be protected by that symmetry, instead it is protected by the bulk $\Theta T_{1/2}$ ($\Theta$)-symmetry-induced topological classification of the ground state.
This property, that the sum of 3D interior and surface contributions 
is in accord with 3D bulk topological constraints even when the separate contributions are nonquantized and nonuniversal,
is reminiscent of the frequently discussed nonuniversal Hall current partition 
between interior and edge \cite{heinonen1985current,thouless1993edge,hirai1994ratio,haremski2020electrically} in the IQHE.

\begin{figure}[t!]
  \centering
  \includegraphics[width=0.9 \linewidth ]{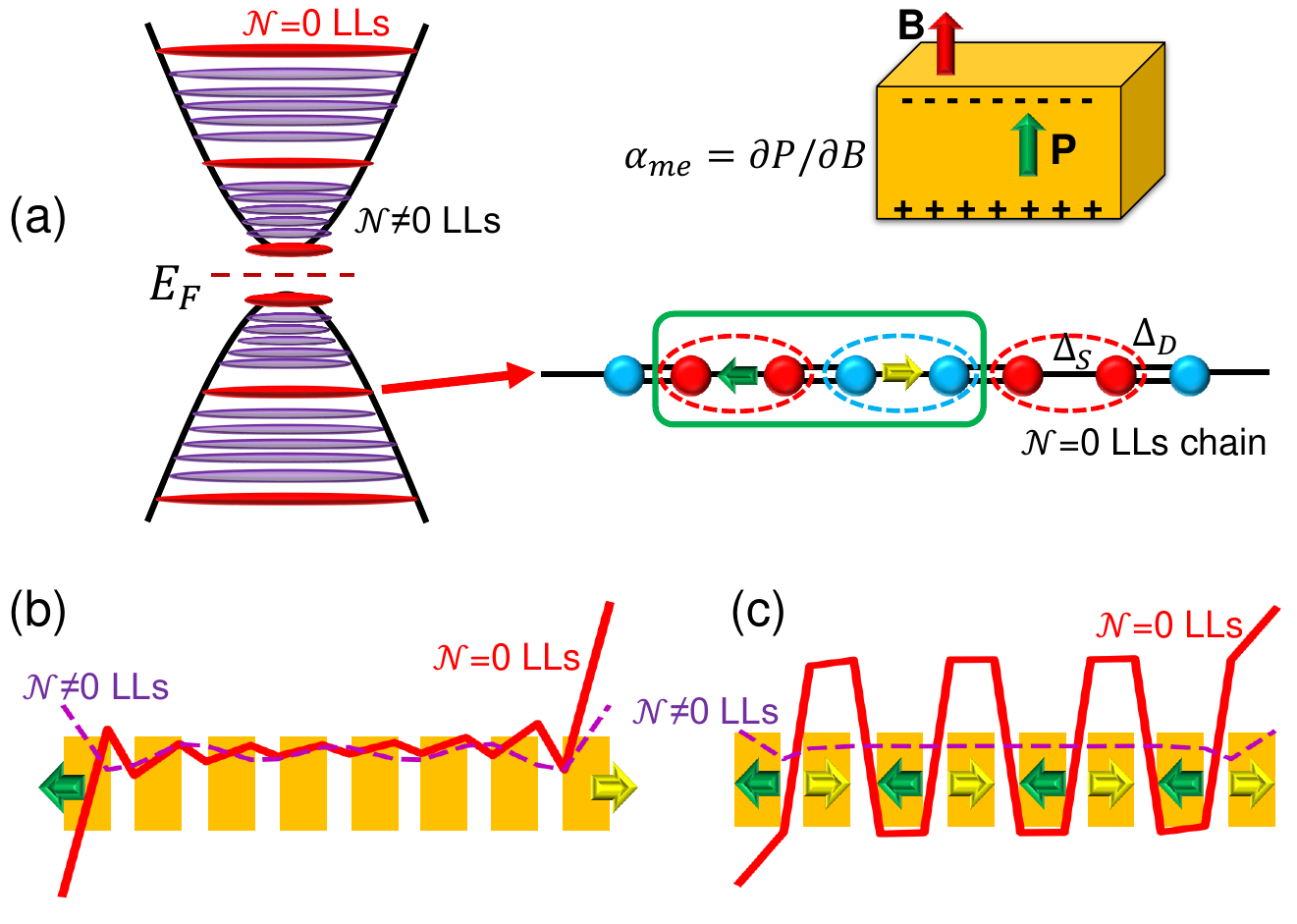}
  \caption{Landau levels (LLs).
  (a) Schematic topological insulator thin film 
  energy spectrum in the absence (black solid) and presence (LL spectra) of a uniform dc magnetic field.
  The low-energy Hamiltonian for an AFM magnetic configuration acts in the ${\cal N} = 0$ LL subspace as a generalized 1D SSH model (see main text) with four electronic states (represented by colored circles) per unit cell (green rectangle), one state at each surface of two magnetic layers (dashed ovals) with opposite magnetization (represented by red and blue colors). 
  The states at each surface of every (interior) layer
  are coupled to the static magnetic moments in two layers that have opposite spin-orientations, 
  one in the same layer with exchange coupling $J_{\text{S}}$ and one in an adjacent layer 
  with exchange coupling $J_{\text{D}}$. Thus, the effective Zeeman energies $\pm (J_{\text{S}}-J_{\text{D}})$ 
  (represented by light and dark green arrows) alternate in sign across the layers. 
  Schematic charge distributions of anomalous ${\cal N} = 0$ LLs (red curves) 
  and ${\cal N} \ne 0 $ LLs (purple dashed curves) in TI thin films: 
  (b) with magnetic dopants only in the exterior layers and (c) 
  with interior magnetic moments and A-type antiferromagnetism.}
  \label{fig:anormalous_LLs}
\end{figure}

Our explicit calculations employ a model that can describe
both nonmagnetic and AFM layered materials, and for each case
can support ground states characterized by a trivial or nontrivial $\mathbb{Z}_{2}$ index.   
The model provides a realistic description of A-type AFM TIs,
similar to the paradigmatic AFM TI MnBi$_2$Te$_4$ \cite{Otrokov_2017,Otrokov2019,Li2019_theory,Deng2020,Chowdhury_2019,Deng_2020,Ge2020,Wimmer2021,MBT_Review_2021,bernevig2022progress}, but is simplified in a way
that is convenient for explicit inclusion of magnetic fields that are parallel to the stacking axis \cite{Lei2020}.
We use a continuum version of this model to calculate electronic charge
distributions across thin films and a lattice regularized version
to address the topological magnetoelectric response of bulk crystals. 
As illustrated in Fig.~\ref{fig:anormalous_LLs}(a), our thin film calculations 
show that the magnetic field induced dipole moment arises
entirely from the model's anomalous $\mathcal{N}=0$ Landau level (LL) subspace, 
the dynamics of which are governed by a generalized Su-Schrieffer-Heeger (SSH) Hamiltonian. 
Because the $\mathcal{N} \neq 0$ LL eigenfunctions are symmetric 
across the thin film, as illustrated in Figs.~\ref{fig:anormalous_LLs}(b) and \ref{fig:anormalous_LLs}(c), they do not contribute to the magnetic field induced dipole moment.  The absence of a dipole moment in the interior of a nonmagnetic film 
is illustrated schematically in Fig.~\ref{fig:anormalous_LLs}(b), and its presence in an AFM film 
in Fig.~\ref{fig:anormalous_LLs}(c).

\section{Anomalous Landau levels}
An effective low-energy Hamiltonian for the electronic states in the layered materials of interest can be constructed \cite{zhang2009topological,Burkov2011,Lei2020} using Dirac cones, one associated with top and bottom surface of each layer,
that are coupled by hybridizing states both within and between layers:
\begin{align}\label{Hamiltonian_dirac}
   \hat{H} = & \sum_{\bk_{\perp},ij,\mu\gamma} \Big[\Big( \, 
   (-1)^i  \hbar v_{D}  (\hat{z} \times \bs) \cdot \bk_{\perp} + m_{i} \sigma_z \Big)_{\mu,\gamma} \delta_{ij}  \nonumber \\
   &\qquad\qquad + \Delta_{ij}(1-\delta_{ij} )\delta_{\mu,\gamma} \Big] \hat{c}_{\bk_{\perp},i,\mu}^{\dagger} \hat{c}_{\bk_{\perp},j,\gamma}.
\end{align}
Here we have chosen $\hat{\bm{z}}$ as the stacking axis, 
$\bk_{\perp}=(k_{x},k_{y})$, $i,j\in\{0,\ldots,2N-1\}$ where $i=2(l-1)$ $(i=2l-1)$ labels states that are associated with the bottom (top) surface of layer $l\in\{1,\ldots,N\}$,
$\mu,\gamma\in\{\uparrow,\downarrow\}$ are spin labels, 
$\Delta_{ij}$ are the intra- and interlayer hybridization parameters,
$v_{_D}$ is the velocity of an isolated Dirac cone, and $m_{i}$ are mass terms 
that account for exchange coupling to static local magnetic moments (which we take to be aligned with the stacking direction).  We will consider only even layer number $N$ 
so that AFM thin films do not have uncompensated moments, and in our model
$m_{i}$ is a mass term at surface $i$ that accounts for exchange coupling to static local magnetic moments. 
We set $m_{i} = \sum_{l} J_{il} M_{l}$ with dimensionless magnetization $M_{l} = 0$ if layer $l$ is nonmagnetic and $\pm 1$ if layer $l$ is magnetic
(for antiferromagnetic configuration we take $M_{l} = (-1)^{l}$);
the Dirac-like states at the top or bottom surface of layer $l$ 
are coupled to the static magnetization of that layer (via $J_\text{S}$)
and to that of the adjacent layer nearest that surface (via $J_\text{D}$),
and hybridized only to adjacent surfaces 
with hopping $\Delta_\text{S}$ for opposite surfaces of the same layer and 
$\Delta_\text{D}$ for the closest surface of the adjacent layer (see Fig.~1 of Ref.~\cite{Lei2020}).
In thin film models the Dirac-like states of the outermost surfaces
do not have $J_\text{D}$ or $\Delta_\text{D}$ coupling.  

We account for an external magnetic field $\bB = -B \hat{\bz}$ though the usual prescription \footnote{See, e.g., Chapter 2 of Winkler \cite{WinklerBook}.} of mapping a $\bk\cdot\bp$ model to an envelope function approximated (real-space) Hamiltonian that is then minimally coupled to the electromagnetic field: 
$\hbar\bk \rightarrow -i\hbar\bm{\nabla}-(eBx/c)\hat{\by}$ in the Landau gauge. 
Here $e>0$ is the elementary charge unit.
As described in Appendix \ref{Appendix:LLs}, 
we apply this recipe to
Eq.~(\ref{Hamiltonian_dirac}) and seek energy eigenfunctions of the form $\Psi_{E,q_{y}}(x,y)=e^{iq_{y}y}\Phi_{E,q_{y}}(x)/\sqrt{L_{y}}$, where 
$\Phi_{E,q_{y}}(x)$ has $2N\times2$ components that 
correspond with layer and spin label degrees of freedom, respectively.
This results in the individual Dirac Hamiltonian $H_{D}$ of each surface being recast in terms of the LL raising and lowering (differential) operators
$a^{\dagger} \equiv (\tilde{x} - \partial_{\tilde{x}})/\sqrt{2}$ and $a \equiv (\tilde{x} + \partial_{\tilde{x}})/\sqrt{2}$,
\begin{equation}
\label{eq:Dirac}
    H_D=\hbar v_{_D}  (\hat{z} \times \bs) \cdot \bk_{\perp} \to \hbar \omega_c \begin{pmatrix} 0 & a^{\dagger} \\ a & 0 \end{pmatrix},
\end{equation}
where
$ \omega_c \equiv \sqrt{2}v_D/l_B $, 
$ l_B \equiv \sqrt{\hbar c/eB} $ is the magnetic length, and $ \tilde{x} \equiv l_B q_{y} - x/l_B $.
Because $H_D$ couples $(\chi_{n}(\tilde{x}),\,0)$ only to $(0,\,\chi_{n-1}(\tilde{x}))$, where $\chi_{n}$ are eigenfunctions of $a^{\dagger}a$ that satisfy $a^{\dagger}\chi_{n}\propto\chi_{n+1}$ and $a\chi_{n+1}\propto\chi_{n}$ for integers $n\ge 0$ \footnote{See, e.g., pg.~89-94 of Ref.~\cite{Sakurai}.},
the total Hamiltonian matrix obtained from Eq.~(\ref{Hamiltonian_dirac}) can be organized into $n_c+1$ blocks of which $n_c$ have 
dimension $2N\times2$ where $n_c$ is a LL index cutoff.
The remaining block contains only anomalous LL states, which have the form $(\chi_{0}(\tilde{x}),\,0)$ at each surface and are spin-up polarized, and has dimension $2N$.
Thus, to each energy eigenfunction we can associate a nonnegative block index ${\cal N}$ (chosen to equal the LL 
index $n$ of its spin-up components) and by an integer $r$ 
\footnote{For ${\cal N}=0$, $r\in\{1,\ldots,2N\}$ and for ${\cal N}>0$, $r\in\{1,\ldots,2N\times 2\}$.}: 
$\Phi_{E,q_{y}}(x)=\Phi_{\mathcal{N}}^{r}(\tilde{x})$, where
$\Phi_{\mathcal{N}}^{r}(\tilde{x})=\prod_{i}\left(C^{{\cal N},r}_{(i,\uparrow)}\chi_{n}(\tilde{x}),\,C^{{\cal N},r}_{(i,\downarrow)}\chi_{n-1}(\tilde{x})\right)$ for $\mathcal{N}>0$ and
$\Phi_{0}^{r}(\tilde{x})=\prod_{i}\left(C^{{0},r}_{(i,\uparrow)}\chi_{0}(\tilde{x}),\,0\right)$.
Within each block all energy eigenvalues occur in pairs that differ only 
in sign. 
The Hamiltonian acts in $\mathcal{N}=0$ block as a finite-size generalized SSH
Hamiltonian (see Appendix \ref{Appendix:LLs1}) illustrated schematically in Fig.~\ref{fig:anormalous_LLs}(a). 
The SSH hopping parameters are $\Delta_\text{S}$ and $\Delta_\text{D}$ and the exchange interaction
($J_\text{S}-J_\text{D}$ on the interior surfaces and $J_\text{S}$ for the top and bottom surface) define SSH model site energies.

Below we calculate the electric dipole moment of finite thickness films by endowing the LL eigenfunctions of the quasi-2D model Eq.~(\ref{Hamiltonian_dirac})
with $z$-dependence by taking $C^{{\cal N},r}_{(i,\mu)}\rightarrow C^{{\cal N},r}_{(i,\mu)}\delta(z-z_{i})$, where $z_{i}\hat{\bz}$ is the location of the $i$th surface 
with the center of the film taken as the origin.
The macroscopic electronic polarization (dipole moment per volume) is:
\begin{equation}
    \bP_{\text{el}} = -\frac{e}{\Omega} \sum_{E < 0} \int \br |\Psi_{E}(\br)|^2 d\br,
    \label{Pfinite}
\end{equation}
where $\Omega=A (z_{2N-1}-z_{0})$ is the volume of the finite sample, $A$ is the film cross sectional area
and we have noted that only negative energy states are occupied at neutrality.
Here the $\Psi_{E}(\br)$ are real-space energy eigenfunctions that have support only within the sample. 
Taking account of the LL degeneracy $A/(2\pi l_B^2)$, we obtain
\begin{align}
    P^{z}_{\text{el}} = -\frac{e^2B}{Ndhc}\sum_{i} \; z_{i} \; \Bigg( \sum_{E_{{\cal N},r}<0,\mu} |C^{{\cal N},r}_{(i,\mu)}|^2 \Bigg)\equiv \alpha_{\text{me}}B^{z}.
\end{align}
The factor in round brackets is the areal electronic charge density on surface $i$ and $d$ is the thickness of the unit cell for each layer.
Since $m_i$ is odd under the mirror transformation $z \to -z$ (see Appendix \ref{Appendix:LLs1}),
it yields a spin-polarization that is odd in $z$.  In the ${\cal N}=0$ anomalous LL subspace, which contains only spin-up states, the mass terms give rise to a charge polarization.  Because the 
block projected Hamiltonian is odd under the combined mirror and spin-reversal operations,
the charge density contributed by the ${\cal N}\ne 0$ blocks is even under $z \to -z$ and there is no
contribution to the dipole moment; the magnetoelectric response of our model 
comes entirely from the ${\cal N}=0$ block.  Since the Hamiltonian of the ${\cal N}=0$ sector
is a generalized SSH model with alternating hopping strengths $\Delta_\text{S}$ and $\Delta_\text{D}$, the 
quantized magnetoelectric coefficient $\alpha_{\text{me}}$ in our model follows from familiar analogous 
polarization properties of the SSH model.  
In the following, we detail how these properties depend on model parameters.

\section{Bulk phase diagram and polarization properties}
\begin{figure}[t!]
  \centering
  \includegraphics[width=0.9 \linewidth ]{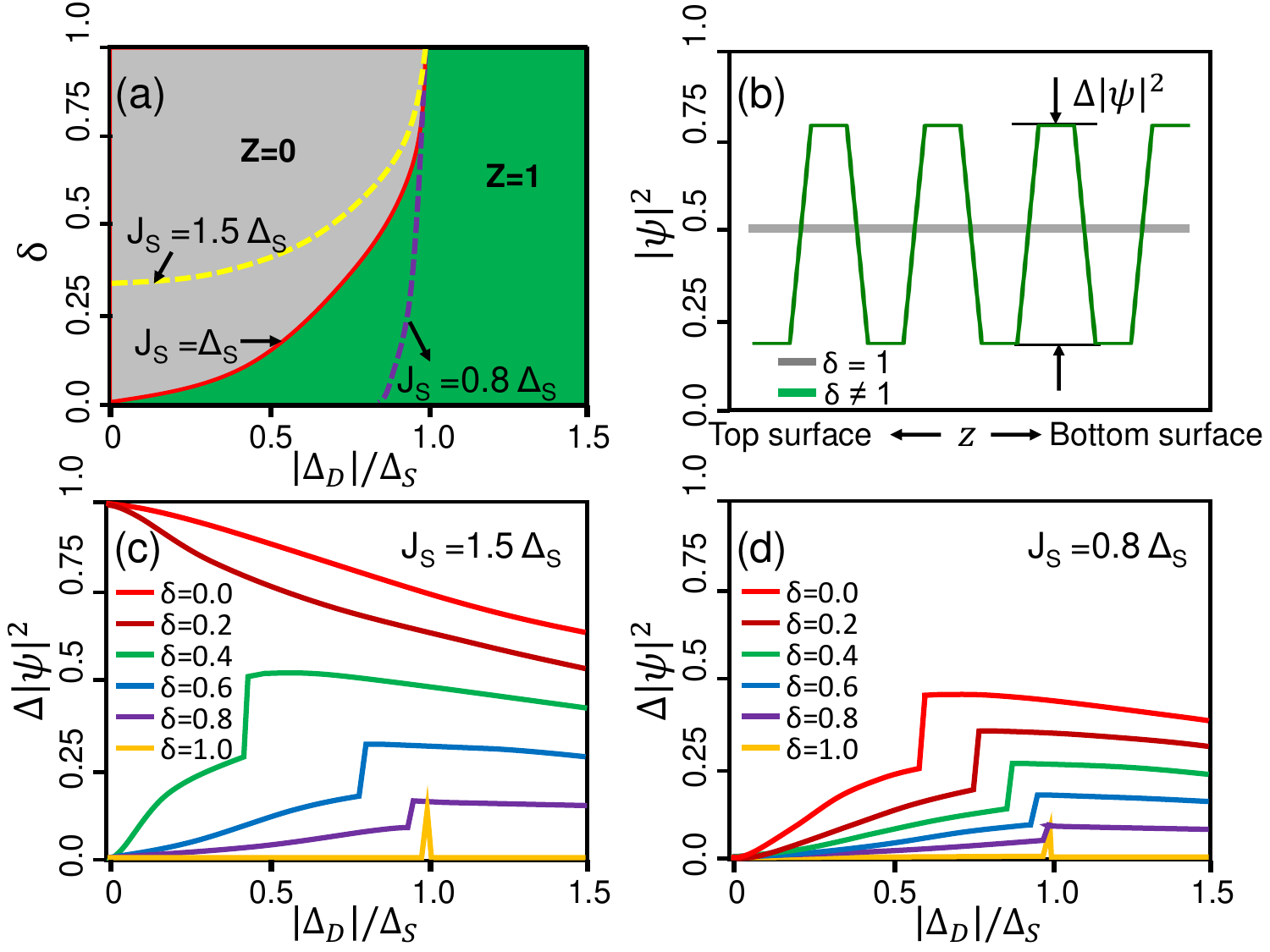}
  \caption{Topological phase diagram and bulk charge redistribution.
  (a) Phase diagram of bulk AFM topological insulators {\it vs.} $\Delta_\text{D}/\Delta_\text{S}$ and $\delta \equiv J_\text{D}/J_\text{S}$ at $\Delta_\text{S}/J_\text{S}=1$.  (The phase 
  boundaries for $J_\text{S} = 0.8 \Delta_\text{S}$ and $J_\text{S} = 1.5 \Delta_\text{S}$ are plotted as 
  purple and yellow dashed lines.) 
  (b) Schematic magnetic-field induced charge density redistribution
  for nonmagnetic (gray curve) and AFM (green curve) TIs;
  (c), (d) Wave-function amplitude variation (defined in panel (b)) {\it vs.} $\Delta_\text{D}$ for various
  values of $\delta$ when $J_\text{S} = 1.5 \Delta_\text{S}$ and $J_\text{S} = 0.8 \Delta_\text{S}$, respectively. 
  }\label{fig:bulk_partition}
\end{figure}

We can obtain the topological phase diagram of the bulk 3D Hamiltonian related to Eq.~(\ref{Hamiltonian_dirac}) by tracking the band gap evolution when periodic boundary conditions are
applied in $\hat{\bz}$ direction 
\footnote{Technically, the model presented in Eq.~(\ref{Hamiltonian_dirac}) is a quasi-2D low-energy effective $\bk\cdot\bm{p}$ Hamiltonian. To generate a genuine 3D model for the electronic states in bulk, we need to replace the $\bk\cdot\bm{p}$ basis states that the quasi-2D Hamiltonian is expressed in terms of with hybrid Wannier functions that are spatially localized in the $\bz$ direction and the set of which contains one spin-up and one spin-down WF localized near each surface. We then apply periodic boundary conditions, by which we mean we take the model to describe an infinite chain of coupled-Dirac cone models (i.e., it involves a sum over all lattice vectors) in the $\bz$-direction.}.
Our findings are shown in Fig.~\ref{fig:bulk_partition}(a).
As explained below, these assignments have been confirmed by calculating the Zak phase of 
a corresponding one-dimensional (1D) model and also 
by explicitly evaluating the topological magnetoelectric coefficient of a 3D lattice regularization of Eq.~(\ref{Hamiltonian_dirac}).
The model parameters are chosen such that $\delta\equiv J_{\text{D}}/J_{\text{S}}\rightarrow 1$ corresponds to the nonmagnetic case,
for which a topological phase transition occurs when $|\Delta_\text{D}| = \Delta_\text{S}$.  In this case there is no 
magnetic-field dependent charge transfer between layers
in the bulk as illustrated by the gray curve in Fig.~\ref{fig:bulk_partition}(b). 
When $\delta \ne 1$ the interior layers are magnetized and magnetic-field dependent charge transfer 
does occur as illustrated by the green curve in Fig.~\ref{fig:bulk_partition}(b). 
Note that in the AFM case the topological phase boundary depends on $J_\text{S}$,
that interlayer charge transfer occurs in both the trivial ($\mathbb{Z}_2$-even) and nontrivial ($\mathbb{Z}_2$-odd) phases with magnitude dependent on model parameters, 
and that the rate of charge transfer with field jumps at the phase transition points, as shown in Figs.~\ref{fig:bulk_partition}(c) and \ref{fig:bulk_partition}(d). 

The magnetic-field dependent charge transfer seems to suggest that the magnetoelectric responses of 
both $\mathbb{Z}_{2}$-even and $\mathbb{Z}_{2}$-odd phases are nonzero, but this is not the case.
As we have described near Eq.~(\ref{eq:Dirac}) and detailed in Appendix \ref{Appendix:LLs}, the low-energy Hamiltonian in the 3D bulk can be block-diagonalized into subspaces that are uniquely identified by a LL index $\mathcal{N}$.
As in finite samples, in this bulk model the magnetoelectric response comes entirely from the  
$\mathcal{N} = 0$ LL subspace, within which the AFM Hamiltonian reduces to a family
of identical 1D generalized SSH models -- 
one for each state in the LL (see Appendix \ref{Appendix:LLs1}): 
\begin{align}
    H_{\text{AFM}}(k_{z}) &= (J_\text{S}-J_\text{D}) \lambda_z\otimes\tau_{0}+ \Delta_\text{S} \lambda_{0}\otimes\tau_x  \nonumber\\
    &+ \Delta_\text{D} \lambda_x\otimes(\tau_x \cos k_z d + \tau_y \sin k_z d ).
    \label{ssh}
\end{align}
In Eq.~(\ref{ssh}) each 1D unit cell has four sites since there are 
two van der Waals layers, each with top and bottom surface Dirac cones, and  
$\lambda_i$ and $\tau_i$ are Pauli matrices that act on the layer and  
surface degrees of freedom.
(Spin degree of freedom does not appear in Eq.~(\ref{ssh}) since states in the $\mathcal{N}=0$ LL subspace are spin-up polarized.)
The ground state polarization in our low-energy 3D model has one bulk component, namely $P^{z}$,
which can be calculated using expressions derived in the modern theory of polarization \cite{Vanderbilt1993,Resta2007}.
The $\mathcal{N}=0$ contribution to $P^{z}$ is proportional to the 1D bulk polarization of Eq.~(\ref{ssh}) at half-filling: $P_{\text{SSH}}^{z}=e\sum_{n=1}^{2}\int_{-\pi/2d}^{\pi/2d}\frac{dk_{z}}{2\pi}\xi^{z}_{nn}(k_{z})$, where 
$\xi^z_{nm}(k_{z})\equiv i\langle u_{n,k_{z}}|\frac{\partial}{\partial k^{z}}u_{m,k_{z}}\rangle$ is the 
Berry connection in a (periodic) Hamiltonian gauge.
$P_{\text{SSH}}^{z}$ is quantized because Eq.~(\ref{ssh}) has
particle-hole symmetry: $U^{\dagger}_{\mathcal{C}}(k_{z})H_{\text{AFM}}(-k_{z})^{*}U_{\mathcal{C}}(k_{z})=-H_{\text{AFM}}(k_{z})$ for $U_{\mathcal{C}}(k_{z})=(\lambda_{x}\text{cos}dk_{z}-\lambda_{y}\text{sin}dk_{z})\otimes\tau_{z}$.
Indeed, at half-filling this model can support topologically distinct ground states
characterized by a $\mathbb{Z}_{2}$-valued invariant \footnote{See Sec.~2D of Ref.~\cite{TME_TI_2008} and Ref.~\cite{ryu2010topological}.}, 
which is obtained by performing the 1D BZ integral in $P^{z}_{\text{SSH}}$.
The association \footnote{See Sec.~2D of Ref.~\cite{TME_TI_2008}.} of $\mathbb{Z}_{2}$-even with $P_{\text{SSH}}^{z} = en$
and $\mathbb{Z}_{2}$-odd with $ P_{\text{SSH}}^{z} = e(n+\frac{1}{2})$ for some $n\in\mathbb{Z}$ 
allows Fig.~\ref{fig:bulk_partition}(a) to be confirmed by a semianalytic calculation (see Appendix \ref{Appendix:LLs2}).
A similar analysis applies in the nonmagnetic case, for which the projected Hamiltonian in the $\mathcal{N} = 0$ subspace is the usual SSH  model.
Thus, $P^{z}$ does not 
distinguish between the microscopically distinct
magnetic field dependent charge redistribution (Fig.~\ref{fig:bulk_partition}) in AFM and nonmagnetic cases.

We can also evaluate the bulk magnetoelectric coefficient in Eq.~(\ref{alpha}) using a 3D tight-binding Hamiltonian \cite{mahon2023reconciling} obtained by regularizing Eq.~(\ref{Hamiltonian_dirac}) and employing
the previously derived \cite{TME_TI_2008,Vanderbilt2009} linear response expression.
As described above, in band insulators with (generalized) TRS it has been argued that $\alpha^{il}=\delta^{il}\alpha_{\text{CS}}$, where $\alpha_{\text{CS}}$ is sensitive to the TRS-induced topological classification of the electronic ground state (encoded in a $\mathbb{Z}_{2}$ index). In general, $\alpha_{\text{CS}}$ is gauge dependent and its explicit form employs a smooth global gauge of the vector bundle of occupied Bloch states over the 3D BZ torus. Thus, a smooth global Hamiltonian gauge $\mathfrak{u}$ identified pointwise by $\mathfrak{u}_{\bm{k}}=(\ket{u_{n\boldsymbol{k}}})_{n\in\{1,\ldots,N_{\text{occ}}\}}$ is viable (if it exists) and in that gauge
\begin{equation}
        \alpha^{\mathfrak{u}}_{\text{CS}}=
        \frac{-e^2}{2\hbar c}\epsilon^{abd}\int_{\text{BZ}} {\frac{d^{3}k}{(2\pi)^3}\left(\xi^a_{vv'}\partial_b\xi^d_{v'v}-\frac{2i}{3}\xi^a_{vv'}\xi^b_{v'v_1}\xi^d_{v_1v}\right)},
        \label{alphaCS}
\end{equation}
where repeated indices are summed.  (The implicit band sums run over the 
occupied bands $v,v',v_1\in\{1,\ldots,N_{\text{occ}}\}$ only.) 
Although the expression for $\alpha^{\mathfrak{u}}_{\text{CS}}$ is gauge-dependent, transforming from one appropriate gauge \footnote{In this context one typically considers gauge transformations that decompose as a product of unitary transformations, one acting on the occupied states and another on the unoccupied states. A gauge choice that is periodic over BZ is always made.} to another can only change its value by an integer multiple of $e^2/hc$ \cite{TME_TI_2008,Vanderbilt2009}.

Explicitly computing $\alpha_{\text{CS}}$ is challenging because a smooth global Hamiltonian gauge does not typically exist in $\mathbb{Z}_{2}$-odd insulators \cite{Vanderbilt2012,Monaco2017}, requiring a nontrivial Wannierization-like procedure to be employed.
In our case, however, because the model has a generalized TRS and all energy bands appear in isolated degenerate pairs over the entire BZ, such a gauge does exist 
\footnote{In general, TRS implies that the (set of) Chern number(s) associated with each set of isolated bands vanishes, thus there exists a smooth global frame of the corresponding vector bundles. Due to the degeneracy in this special case, the components of each such frame are energy eigenvectors. (This same argument holds if there are no degeneracies between bands anywhere in BZ, but in the presence of a symmetry such as time-reversal or inversion, which ultimately gives rise to a $\mathbb{Z}_{2}$ index, this will never be the case.) Hence, a smooth global Hamiltonian gauge always exists in such circumstances.}. 
In a related work \cite{mahon2023reconciling} we explicitly evaluated Eq.~(\ref{alphaCS}) in the nonmagnetic case. We can, at the cost of some additional complication, also find 
the analytic form of a smooth global Hamiltonian gauge in the AFM case. 
With this gauge in-hand, and employing the previously detailed \cite{mahon2023reconciling} approximation that the basis of the tight-binding model (that results from lattice regularization) 
consists of atomic-like Wannier functions, we can semianalytically evaluate Eq.~(\ref{alphaCS}). 
(See Appendix \ref{Appendix:3DTightBinding1} and \ref{Appendix:3DTightBinding2}.) 
This calculation again reproduces quantized $\alpha_{\text{CS}}$ assignments in agreement with
Fig.~\ref{fig:bulk_partition}(a).

\section{Magnetoelectric response in thin films} 
\begin{figure}[t!]
  \centering
  \includegraphics[width=0.9 \linewidth ]{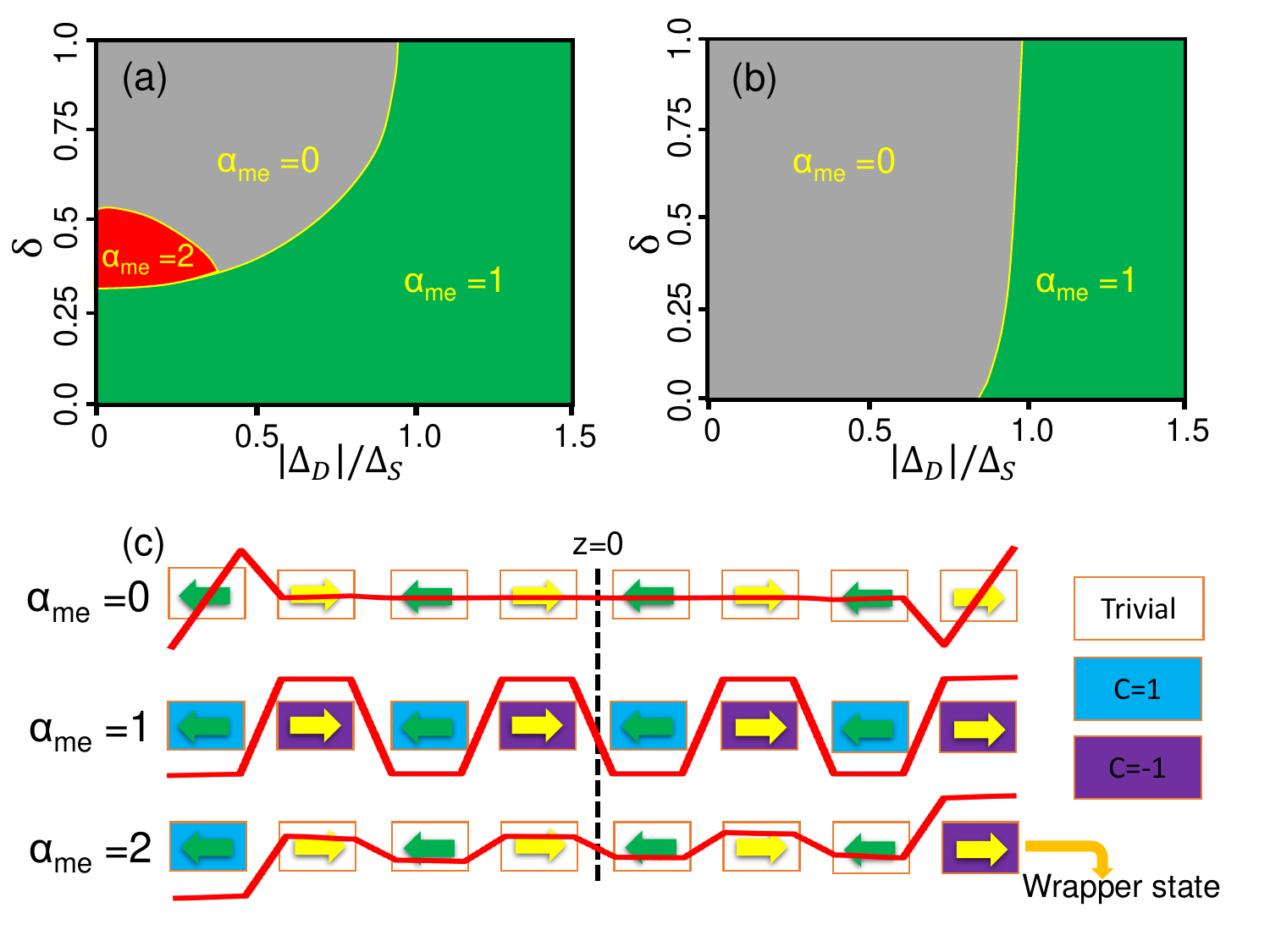}
  \caption{
  (a) Thin film phase diagram for $J_\text{S} =1.5 \Delta_\text{S}$. The (quantized) magnetoelectric coefficient $\alpha_{\text{me}}$ is expressed here in units of $e^2/2hc$. 
  The green region corresponds to $\mathbb{Z}_2$-odd and the red and gray regions to $\mathbb{Z}_2$-even bulk phases; the states in the red region have Chern wrappers as discussed in the main text, and therefore have $\alpha_{\text{me}} = 2$. 
  (b) Thin film phase diagram for $J_\text{S} =0.8 \Delta_\text{S}$.
  (c) Schematic illustration of ${\cal N}=0$ (see main text) 
  charge distributions (red curves) for $\alpha_{\text{me}} = 0,\,1,\,2$ at $J_\text{S} =1.5 \Delta_\text{S}$ and a 
  small value of $\Delta_\text{D}$.  In all cases there are both interior and surface contributions to $\alpha_{\text{me}}$ that are not quantized separately.  The arrows indicate the orientation of the static magnetic moments in each van der Waals layer and the background color (white, blue, green) 
  indicates the layer Chern number in the limit $\Delta_\text{D} \to 0$.
  }\label{fig:ME_thin_films}
\end{figure}

In Fig.~\ref{fig:ME_thin_films}(a) we show a 
$J_\text{S}=1.5 \Delta_\text{S}$ phase diagram constructed by 
tracking thin film gap closings and confirmed by
evaluating the magnetoelectric response coefficient $\alpha_{\text{me}}$,
expressed here in units of $e^2/2hc$.
The thin-film results perfectly match the bulk $J_\text{S}=1.5 \Delta_\text{S}$ results in Fig.~\ref{fig:bulk_partition}.
Notably, the thin-film phase diagram has an extra boundary surrounding the region with $\alpha_{\text{me}}=2$, which results 
from the appearance of (top and bottom) surface Hall conductivities (Chern wrappers \cite{vanderbilt_book})
in this region of our model's phase diagram, that are not related to bulk topology.
The Chern wrapper region's extent depends on details of the model that 
do not influence bulk topology; it is
absent for $J_\text{S} \le \Delta_\text{S}$, as illustrated in Fig.~\ref{fig:ME_thin_films}(b).  
Note that the opportunity to identify bulk topological phase transitions by tracking thin film
gaps is present only in magnetic $\mathbb{Z}_{2}$ TIs, which do not have gapless edge modes.

Figures \ref{fig:ME_thin_films}(a) and \ref{fig:ME_thin_films}(b) can be understood qualitatively by 
examining the $\Delta_\text{D}=0$ limit of the ${\cal N}=0$ Hamiltonian.
In this hypothetical limit there is no hybridization between van der Waals layers and 
each layer is described by independent $2 \times 2$ Hamiltonians with
a surface degree of freedom:
\begin{equation}
    H_{\text{1L}} = \bar{m} \tau_0 + \tilde{m} \tau_z + \Delta_\text{S} \tau_x ,
\end{equation}
where $\bar m \equiv (m_t + m_b)/2$ and $\tilde{m} \equiv (m_t - m_b)/2$ with $m_{t/b}$ the exchange coupling at the top and bottom surface of each layer. On the outermost surfaces $m_{t/b} = \pm J_\text{S}$ whereas on the interior surfaces $m_{t/b} = \pm (1-\delta)J_\text{S}$, with the sign specified by the moment orientation in 
the same layer (see Appendix \ref{Appendix:LLs3}).  The eigenvalues of this Hamiltonian are $E = \bar{m} \pm (\Delta_\text{S}^2 + \tilde{m}^2)^{1/2}$.  The phase diagram is controlled by the relative sizes of 
$|\bar{m}|$ and $(\Delta_\text{S}^2 + \tilde{m}^2)^{1/2}$ on the outermost and interior layers.
In the $\alpha_{\text{me}}=0$ phase, $|\bar{m}| < (\Delta_\text{S}^2 + \tilde{m}^2)^{1/2}$ in all layers, in the $\alpha_{\text{me}}=1$ phase $|\bar{m}| > (\Delta_\text{S}^2 + \tilde{m}^2)^{1/2}$ in all layers, whereas  
in the $\alpha_{\text{me}}=2$ phase $|\bar{m}|$ is larger in the outside layers and 
smaller in the interior layers.
In the $\alpha_{\text{me}}=1$ case, the ${\cal N}=0$ 
eigenfunctions occupy alternate layers as illustrated in the middle
panel of Fig.~\ref{fig:ME_thin_films}(c).  For $\delta=0$ all layers are identical and 
\begin{equation}
    \alpha_{\text{me}} = \sum_{j=1}^{N/2} (-1)^{N/2+j} \frac{2j-1}{N/2}=1
\end{equation}
is quantized even at finite $N$.  More generally, for any values of $\Delta_\text{D}$ and $\delta$,
the magnetoelectric coefficient achieves quantization only in sufficiently thick films (see Appendix \ref{Appendix:LLs3}).

\section{Discussion}
We have explored the magnetoelectric properties of layered
$\mathbb{Z}_{2}$ topological insulators by evaluating the (electronic) polarization response
of realistic thin film and bulk model Hamiltonians to a uniform dc magnetic field along the stacking axis. 
For an effective low-energy model we find that the polarization response arises entirely from 
the formation of $\mathcal{N}=0$ anomalous Landau levels described by 1D
generalized SSH models.
In the thin film case the total magnetoelectric response is 
quantized for sufficiently thick films, but it arises from nonzero and nonuniversal surface and interior contributions.
(This behavior is partially related to that of the alternating stacked Haldane model considered in Ref.~\cite{Souza2017}.)
We have reached these conclusions both by numerically examining the many-layer 
limit of the thin film calculations and by applying periodic boundary conditions in the stacking direction to the SSH models.
Both approaches find that only the total polarization is quantized, 
in agreement with our 3D bulk calculations that relate the magnetoelectric response to the Chern-Simons 3-form \cite{TME_TI_2008,Vanderbilt2009}, which we evaluate explicitly for a lattice regularization of the low-energy model.

It is interesting to relate our results to efforts \cite{Resta2005,Resta2006} to express the the orbital 
magnetization $\bM$ as a sum of contributions
having either atomic-like ($\overline{\bM}$) or itinerant ($\widetilde{\bM}$) character.
This partitioning can be made unambiguously, given a model with isolated energy bands that support 
localized Wannier functions, and associates $\overline{\bM}$ with interior currents 
and $\widetilde{\bM}$ with surface currents.  However, when the bands are not isolated the 
partitioning is gauge-dependent.  In previous work \cite{mahon2023reconciling} on the nonmagnetic case we 
were able, thanks to the analytic nature of our simplified model calculations, to demonstrate that 
there exists a smooth global (Hamiltonian) gauge $\mathfrak{u}$ in which the topological 
magnetoelectric response is entirely itinerant, \textit{i.e.}~$\widetilde{\bM}_{\mathfrak{u}}^{(E)}=\alpha^{\mathfrak{u}}_{\text{CS}}\bE$. 
We conjectured that this property is not specific to the simplified model, but that it is
instead a bulk manifestation of the feature that in nonmagnetic thin films magnetoelectric response can occur only near the surface,
and as a bulk harbinger of the magnetic-dopant requirement for the topological magnetoelectric effect to manifest in nonmagnetic TI films. 
In Appendix \ref{Appendix:3DTightBinding3} we explicitly demonstrate that in the AFM case there \textit{does not} generically exist a smooth global gauge in which $\widetilde{\bM}_{\mathfrak{u}}^{(E)}$ equals $\alpha^{\mathfrak{u}}_{\text{CS}}\bE$.
This difference between AFM and nonmagnetic cases,
which is encoded in the geometry
\footnote{Recall that each global trivialization of a vector bundle corresponds with a global smooth frame (i.e., gauge) thereof (see, e.g., Chapter 10 of Ref.~\cite{LeeBook}). And although, for a given band insulator, the nonexistence of certain smooth global gauges
has been related \cite{Monaco2017} to the topology of the vector bundle of occupied Bloch states $\mathcal{V}\rightarrow\text{BZ}$ \cite{monaco2015symmetry}, we do
not suspect this to be the case here; we anticipate that in an
AFM TI there could exist some smooth global gauges $\mathfrak{u}$ for
which there exists a finely tuned point in the parameter space
where $\bM^{(E)}_{\mathfrak{u}}$ equals $\alpha_{\text{CS}}^{\mathfrak{u}}\bE$. Thus, whether such a gauge generically exists is taken to be a geometric property of $\mathcal{V}$.} 
of the corresponding vector bundles of occupied Bloch states,
can be understood as a bulk manifestation of magnetoelectric response occurring both in the interior and at the surface of AFM TIs.
Thus, although $\alpha_{\text{CS}}$ is determined by the (generalized) TRS-induced $\mathbb{Z}_{2}$
topological classification of the electronic ground state, which does not distinguish between nonmagnetic and AFM TIs, 
the nature of that response -- whether it is entirely itinerant or not -- does distinguish them.

Our results provide a conditional answer to the thorny conceptual issue
as to whether the topological magnetoelectric effect should be viewed as an interior or surface effect.
We have previously emphasized that the magnetoelectric response of ordinary nonmagnetic 
TIs is localized at their surfaces, and is realized only when the surfaces are 
gapped by local magnetic dopants.  When the bulk is magnetic, surface dopants are no longer required
and a response in the interior is allowed by symmetry.  Our finding that the interior response does 
occur but is not (separately) quantized, shows that it is the total response that is related to bulk topology,
and that the mechanism by which it is manifested generically involves both interior and surface contributions.
Indeed, we expect that this periodic microscopic scale magnetic-field-induced charge density variation that appears within AFM TIs (and is absent in nonmagnetic TIs) can be observed using high-sensitivity high-resolution x-ray imaging techniques \cite{disa2020,deumel2021,Wakabayashi2010,morisaki2014,bjorck2008,kumah2013}, and that these measurements might be aided by the fact that the density varies only in the magnetic field direction.
The inability to assign a quantized response of a topological state wholly to interior or surface
is analogous to the related issue of nonuniversal interior-edge current partition \cite{heinonen1985current,thouless1993edge,hirai1994ratio,haremski2020electrically} 
in the IQHE.  In both cases, bulk topology constrains the total response but not its 
partitioning.  

\section{Acknowledgements}
Work at the University of Texas at Austin was supported by the
Robert A. Welch Foundation under Grant Welch F-2112.

\appendix

\begin{widetext}
\section{Landau level representation}
\label{Appendix:LLs}
A previously developed \cite{Lei2020} coupled-Dirac cone model is used to describe the low-energy electronic states in the layered compounds that are of interest in this paper.
This model retains two Dirac cones in 
each layer as low-energy electronic degrees of freedom and the total effective Hamiltonian is
\begin{equation}\label{supp:Hamiltonian_dirac}
   \hat{H} =  \sum_{\bk_{\perp},ij,\mu\gamma} \Big[\Big( \, 
   (-1)^i  \hbar v_{D}  (\hat{z} \times \bs) \cdot \bk_{\perp} + m_{i} \sigma_z \Big)_{\mu,\gamma} \delta_{ij} 
    + \Delta_{ij}(1-\delta_{ij} )\delta_{\mu,\gamma} \Big] \hat{c}_{\bk_{\perp},i,\mu}^{\dagger} \hat{c}_{\bk_{\perp},j,\gamma}.
\end{equation}
Here $\bk_{\perp}\equiv(k_{x},k_{y},0)$ for stacking axis $\hat{\bm{z}}$, $i,j\in\{0,\ldots,2N-1\}$ where $i=2(l-1)$ $(i=2l-1)$ label states that are associated with the bottom (top) surface of layer $l\in\{1,\ldots,N\}$ and
$\mu,\gamma\in\{\uparrow,\downarrow\}$ label up and down spin components of $\sigma_{z}$,
$\Delta_{ij}$ are intra- and interlayer hybridization parameters,
$v_{_D}$ is the velocity of the isolated Dirac cones, and $m_{i}$ are mass terms that account for exchange coupling of the Dirac-like states at surface $i$ to static local magnetic moments (taken aligned with the stacking direction) that can be present in each layer.  We set $m_{i} = \sum_{l} J_{il} M_{l}$ with dimensionless magnetization $M_{l} = 0$ if layer $l$ is nonmagnetic layers and $\pm 1$ if layer $l$ is magnetic. 
As mentioned in the main text, we employ the usual prescription to account for an external magnetic field $\bB = -B \hat{\bz}$ in a $\bk\cdot\bm{p}$ model: 
$\hbar\bk_{\perp} \rightarrow -i\hbar(\partial/\partial x,\,\partial/\partial y,0)$ and choose the Landau gauge $ \bA = (0,-Bx,0) $.
Under this recipe the Dirac Hamiltonian for each surface transforms as $H_D(\bk_{\perp})= \hbar v_D (\sigma_x k_y - \sigma_y k_x) \rightarrow H_{D}(x,y)\equiv\hbar v_D \big(-\sigma_x(i\frac{\partial}{\partial y} + x/l_B^2) + i \sigma_y\frac{\partial}{\partial x}\big)$ with $ l_B \equiv \sqrt{\hbar c/eB}$.
The total Hamiltonian that is obtained from employing this procedure in Eq.~(\ref{supp:Hamiltonian_dirac}) has translational symmetry in $\hat{\by}$, thus energy eigenfunctions can be written as
\begin{equation}
\Psi_{E,q_y}(x,y) = \frac{e^{iq_y y}}{\sqrt{L_{y}}} \Phi_{E,q_y}(x)\equiv
\frac{e^{iq_y y}}{\sqrt{L_{y}}}\prod_{i=0}^{2N-1} \Big(\phi_{E,q_y,(i,\uparrow)}(x),\,
    \phi_{E,q_y,(i,\downarrow)}(x)\Big).
\end{equation}

Since the only position dependence of the total Hamiltonian appears via the terms resembling $H_D$ for each surface, we focus on those contributions first.
For an isolated surface of a nonmagnetic layer the energy eigenfunctions satisfy $H_D^{2}(x,y) \Psi_{E,q_y}(x,y) = E^{2} \Psi_{E,q_y}(x,y)$ and $\Phi_{E,q_y}(x)
= \big(\phi_{E,q_y,\uparrow}(x),\,\phi_{E,q_y,\downarrow}(x)\big)$, where
\begin{align}
(\hbar v_D)^2 (\partial_x +q_y-x/l_B^2) (-\partial_x +q_y-x/l_B^2) \phi_{E,q_y,\uparrow}(x) &= E^2 \phi_{E,q_y,\uparrow}(x),\nonumber\\
(\hbar v_D)^2 (-\partial_x +q_y-x/l_B^2) (\partial_x +q_y-x/l_B^2)  \phi_{E,q_y,\downarrow}(x) &= E^2 \phi_{E,q_y,\downarrow}(x).
\label{eigensys}
\end{align}
With the substitutions $ \tilde{x} \equiv l_B q_y - x/l_B $ and $ \partial _{\tilde{x}} = -l_B \partial_x $, 
define the usual harmonic oscillator creation and annihilation differential operators $a^{\dagger} \equiv (\tilde{x} - \partial_{\tilde{x}})/\sqrt{2}$ and $a \equiv (\tilde{x} + \partial_{\tilde{x}})/\sqrt{2}$, which satisfy $[a,a^{\dagger}]=1$. Then, Eq.~(\ref{eigensys}) is recast as
\begin{equation}
2\left(\frac{\hbar v_D}{l_B}\right)^2 \begin{pmatrix} 
a^{\dagger}a & 0\\
0 & a a^{\dagger}
\end{pmatrix}\Phi_{E,q_y}(x)
\equiv H^{2}_{D}(x,\partial_x;q_{y})\Phi_{E,q_y}(x)=E^{2}\Phi_{E,q_y}(x).
\label{eigensys1}
\end{equation} 
From the usual analysis of the harmonic oscillator \footnote{See, e.g., pg.~89-94 of Ref.~\cite{Sakurai}.} we have that eigenfunctions of $a^{\dagger}a$ are
\begin{equation}
    \chi_{n} (\tilde{x}) = H_n(x-q_y l_B^2) e^{-(x-q_yl_B^2)^2/4l_B^2},
\end{equation}
where $H_n(x)$ is the Hermite polynomial $H_n(x) = (-1)^ne^{x^2} \frac{d^n}{dx^n}e^{-x^2}$, and have corresponding eigenvalues $n$.
These functions satisfy
$a^{\dagger}\chi_{n}(\tilde{x})\propto\chi_{n+1}(\tilde{x})$ and $a\chi_{n+1}(\tilde{x})\propto\chi_{n}(\tilde{x})$ for integers $n\ge 0$.
Using this we find that solutions of the eigenfunction-eigenvalue equation (\ref{eigensys1}) can be written in the general form
\begin{equation}
\Phi_{E,q_{y}}(x)\equiv\Phi_{n}(\tilde{x}) = \begin{cases}
\big(C^{n}_{\uparrow}\chi_{n}(\tilde{x}),\, C^{n}_{\downarrow} \chi_{n-1}(\tilde{x})\big), &\text{ for } n\neq 0\\
\big(C^{0}_{\uparrow}\chi_{0}(\tilde{x}),\, 0\big), &\text{ for } n=0
\end{cases}
\end{equation}
for $C^{n}_{\mu}\in\mathbb{C}$ and the corresponding eigenvalues are $E_n^2 = 2n(\hbar v_D/l_B)^2$. 
This implies $E_n = \pm \hbar\omega_{c}\sqrt{n}$ where $\omega_c \equiv \sqrt{2}v_D/l_B$. 
Indeed, these general forms can diagonalize the isolated nonmagnetic surface Dirac Hamiltonian
\begin{equation}
H_D(\bk_{\perp}) \rightarrow H_{D}(x,\partial_x;q_{y})\equiv \hbar\omega_{c} \begin{pmatrix}
 0 & a^{\dagger} \\
 a & 0
 \end{pmatrix}.
\end{equation}

In the total Hamiltonian (for a $N$ layer film) there is a contribution related to $H_D$ for each surface $i$ that couples $(\chi_{n}(\tilde{x}),0)$ only to $(0,\chi_{n-1}(\tilde{x}))$. 
A reasonable ansatz for an energy eigenfunction could therefore involve $\Phi_{E}(\tilde{x})$ of the form
$\prod_{i=0}^{2N-1} \big(C^{n_{i}}_{(i,\uparrow)}\chi_{n_{i}}(\tilde{x}),\, C^{n_{i}}_{(i,\downarrow)} \chi_{n_{i}-1}(\tilde{x})\big)$,
where $n_{i}\ge 0$ for all surfaces $i\in\{1,\ldots,2N-1\}$.
However, since there are no other terms in Eq.~(\ref{supp:Hamiltonian_dirac}) that involve $\bk_{\perp}$, there are no other terms in the Hamiltonian (after $\hbar\bk_{\perp} \rightarrow -i\hbar(\partial/\partial x,\,\partial/\partial y)$) that involve $a$ or $a^{\dagger}$.
That is, when acting on the ansatz eigenfunction the other terms in the Hamiltonian can map a component at surface $i$ into a component at surface $j$, but only in such a way that the new component at surface $j$ is in the space spanned by $\chi_{n_{i}}(\tilde{x})$.
In our model the states at each surface are coupled to those at two surfaces adjacent to it (via $\Delta_{\text{S}}$ and $\Delta_{\text{D}}$), thus the ansatz is valid only if $n_{1}=\ldots=n_{2N-1}\equiv n$ for $n\ge 0$.
(We sometimes adopt the notation $\mathcal{N}$ for this $n$.)
The total energy eigenfunctions can therefore be written in the general form
\begin{equation}
\Psi_{n,q_{y}}(x,y) = \begin{cases}
\frac{e^{iq_{y}y}}{\sqrt{L_{y}}}\prod_{i=0}^{2N-1} \big(C^{n}_{(i,\uparrow)}\chi_{n}(\tilde{x}),\, C^{n}_{(i,\downarrow)} \chi_{n-1}(\tilde{x})\big), &\text{ for } n\neq 0\\
\frac{e^{iq_{y}y}}{\sqrt{L_{y}}}\prod_{i=0}^{2N-1} \big(C^{0}_{(i,\uparrow)}\chi_{0}(\tilde{x}),\, 0\big), &\text{ for } n=0
\end{cases}.
\label{LLbasis}
\end{equation}

\subsection{Multi-layer antiferromagnetic materials}
\label{Appendix:LLs1}
In this section we specify the action of the Hamiltonian (for a material consisting of $N$ layers) in the subspace defined by LL index $n$ via a matrix acting on a collection of eigenfunction components $C^{n}_{(i,\mu)}$ introduced in Eq.~(\ref{LLbasis}). 
For $n\neq 0$ the Hamiltonian matrix acts on the collection of components $\prod_{i=0}^{2N-1} \big(C^{n}_{(i,\uparrow)},\, C^{n}_{(i,\downarrow)}\big)$.
The bulk ($N\rightarrow\infty$) multi-layer antiferromagnetic Hamiltonian matrix for $n\ne 0$ is
\begin{equation}
  H_{\text{bulk}}^n = \left(
      \begin{array}{cccccccc}
        J_{\text{S}}-J_{\text{D}} & \sqrt{n} \hbar \omega_c & \Delta_{\text{S}} & 0 & 0 & 0 & \Delta_{\text{D}} e^{-2ik_zd} & 0 \\
        \sqrt{n} \hbar \omega_c & -J_{\text{S}}+J_{\text{D}} & 0 & \Delta_{\text{S}} & 0 & 0 & 0 & \Delta_{\text{D}} e^{-2ik_zd} \\ 
        \Delta_{\text{S}} & 0 & J_{\text{S}}-J_{\text{D}} & \sqrt{n} \hbar \omega_c & \Delta_{\text{D}} & 0 & 0 & 0 \\
        0 & \Delta_{\text{S}} & \sqrt{n} \hbar \omega_c & -J_{\text{S}}+J_{\text{D}} & 0 & \Delta_{\text{D}} & 0 & 0 \\ 
        0 & 0 & \Delta_{\text{D}} & 0 & -J_{\text{S}}+J_{\text{D}} & \sqrt{n} \hbar \omega_c & \Delta_{\text{S}} & 0 \\
        0 & 0 & 0 & \Delta_{\text{D}} & \sqrt{n} \hbar \omega_c & J_{\text{S}}-J_{\text{D}} & 0 & \Delta_{\text{S}} \\ 
        \Delta_{\text{D}} e^{2ik_zd} & 0 & 0 & 0 & \Delta_{\text{S}} & 0 & -J_{\text{S}}+J_{\text{D}} & \sqrt{n} \hbar \omega_c \\
        0 & \Delta_{\text{D}} e^{2ik_zd} & 0 & 0 & 0 & \Delta_{\text{S}} & \sqrt{n} \hbar \omega_c & J_{\text{S}}-J_{\text{D}} \\
      \end{array}
      \right).
      \label{Hdirac_AFM}
\end{equation}
For a thin film, for example, a film consisting of two layers ($N=2$), the Hamiltonian matrix is
\begin{equation}
  H_{\text{2L}}^n = \left(
      \begin{array}{cccccccc}
        J_{\text{S}} & \sqrt{n} \hbar \omega_c & \Delta_{\text{S}} & 0 & 0 & 0 & 0 & 0 \\
        \sqrt{n} \hbar \omega_c & -J_{\text{S}} & 0 & \Delta_{\text{S}} & 0 & 0 & 0 & 0 \\ 
        \Delta_{\text{S}} & 0 & J_{\text{S}}-J_{\text{D}} & \sqrt{n} \hbar \omega_c & \Delta_{\text{D}} & 0 & 0 & 0 \\
        0 & \Delta_{\text{S}} & \sqrt{n} \hbar \omega_c & -J_{\text{S}}+J_{\text{D}} & 0 & \Delta_{\text{D}} & 0 & 0 \\ 
        0 & 0 & \Delta_{\text{D}} & 0 & -J_{\text{S}}+J_{\text{D}} & \sqrt{n} \hbar \omega_c & \Delta_{\text{S}} & 0 \\
        0 & 0 & 0 & \Delta_{\text{D}} & \sqrt{n} \hbar \omega_c & J_{\text{S}}-J_{\text{D}} & 0 & \Delta_{\text{S}} \\ 
        0 & 0 & 0 & 0 & \Delta_{\text{S}} & 0 & -J_{\text{S}} & \sqrt{n} \hbar \omega_c \\
        0 & 0 & 0 & 0 & 0 & \Delta_{\text{S}} & \sqrt{n} \hbar \omega_c & J_{\text{S}} \\
      \end{array}
      \right).
\end{equation}
Under the mirror transformation $z\rightarrow -z$, which we encode in a unitary operator $M$ that maps the spin-$\sigma$ orbital at the surface located at $z_{i}\hat{\bz}$ to the spin-$\sigma$ orbital located at $-z_{i}\hat{\bz}$ (thus, $M=M^{-1}$),
we find
\begin{equation}
  MH_{\text{2L}}^nM = \left(
      \begin{array}{cccccccc}
        -J_{\text{S}} & \sqrt{n} \hbar \omega_c & \Delta_{\text{S}} & 0 & 0 & 0 & 0 & 0 \\
        \sqrt{n} \hbar \omega_c & J_{\text{S}} & 0 & \Delta_{\text{S}} & 0 & 0 & 0 & 0 \\ 
        \Delta_{\text{S}} & 0 & -J_{\text{S}}+J_{\text{D}} & \sqrt{n} \hbar \omega_c & \Delta_{\text{D}} & 0 & 0 & 0 \\
        0 & \Delta_{\text{S}} & \sqrt{n} \hbar \omega_c & J_{\text{S}}-J_{\text{D}} & 0 & \Delta_{\text{D}} & 0 & 0 \\ 
        0 & 0 & \Delta_{\text{D}} & 0 & J_{\text{S}}-J_{\text{D}} & \sqrt{n} \hbar \omega_c & \Delta_{\text{S}} & 0 \\
        0 & 0 & 0 & \Delta_{\text{D}} & \sqrt{n} \hbar \omega_c & -J_{\text{S}}+J_{\text{D}} & 0 & \Delta_{\text{S}} \\ 
        0 & 0 & 0 & 0 & \Delta_{\text{S}} & 0 & J_{\text{S}} & \sqrt{n} \hbar \omega_c \\
        0 & 0 & 0 & 0 & 0 & \Delta_{\text{S}} & \sqrt{n} \hbar \omega_c & -J_{\text{S}} \\
      \end{array}
      \right).
\end{equation}
When composed with a spin-flip operation each $n\neq 0$ Hamiltonian is invariant, thus the spin-flip times mirror operator is a symmetry.

We can similarly consider the multi-layer antiferromagnetic Hamiltonian for $n=0$, in which case the matrix acts on the collection of components $\prod_{i=0}^{2N-1} \big(C^{n}_{(i,\uparrow)}\big)$. 
In the bulk case, the Hamiltonian matrix is
\begin{equation}
  H^{n=0}_{\text{bulk}} = \left(
      \begin{array}{cccc}
        J_{\text{S}}-J_{\text{D}} &  \Delta_{\text{S}} & 0  & \Delta_{\text{D}} e^{-2ik_zd} \\
        \Delta_{\text{S}} & J_{\text{S}}-J_{\text{D}} & \Delta_{\text{D}} & 0 \\
        0 & \Delta_{\text{D}} & -J_{\text{S}}+J_{\text{D}} & \Delta_{\text{S}}  \\
        \Delta_{\text{D}} e^{2ik_zd} & 0 & \Delta_{\text{S}} & -J_{\text{S}}+J_{\text{D}} \\
      \end{array}
      \right).
      \label{Hamiltonian_N=0}
\end{equation}
In the case of a two-layer thin film the Hamiltonian matrix is 
\begin{equation}
  H^{n=0}_{\text{2L}} = \left(
      \begin{array}{cccc}
        J_{\text{S}} &  \Delta_{\text{S}} & 0  & 0 \\
        \Delta_{\text{S}} & J_{\text{S}}-J_{\text{D}} & \Delta_{\text{D}} & 0 \\
        0 & \Delta_{\text{D}} & -J_{\text{S}}+J_{\text{D}} & \Delta_{\text{S}}  \\
        0 & 0 & \Delta_{\text{S}} & -J_{\text{S}} \\
      \end{array}
      \right).
      \label{Hamiltonian_N=0_film}
\end{equation}
In contrast to the $n\neq0$ case, $H^{n=0}_{\text{2L}}$ is not symmetric under the spin-flip times mirror transformation.

At $k_{z}=0$ the energy eigenvalues of $H^{n=0}_{\text{bulk}}$ (Eq.~(\ref{Hamiltonian_N=0})) are:
\begin{align}
E_1 &= -E_4 = \sqrt{\Delta_{\text{D}}^2 + (J_{\text{S}}-J_{\text{D}})^2} + \Delta_{\text{S}},\nonumber\\
E_2 &= -E_3 = \sqrt{\Delta_{\text{D}}^2 + (J_{\text{S}}-J_{\text{D}})^2} - \Delta_{\text{S}},
\end{align}
with corresponding (nonnormalized) eigenvectors:
\begin{align}
\Psi_1 &= \left( \frac{J_{\text{S}}-J_{\text{D}} + \sqrt{\Delta_{\text{D}}^2 + (J_{\text{S}}-J_{\text{D}})^2}}{\Delta_{\text{D}}}, \frac{J_{\text{S}}-J_{\text{D}} + \sqrt{\Delta_{\text{D}}^2 + (J_{\text{S}}-J_{\text{D}})^2}}{\Delta_{\text{D}}}, 1, 1 \right),\nonumber\\
\Psi_2 &= \left( \frac{J_{\text{S}}-J_{\text{D}} + \sqrt{\Delta_{\text{D}}^2 + (J_{\text{S}}-J_{\text{D}})^2}}{\Delta_{\text{D}}}, -\frac{J_{\text{S}}-J_{\text{D}} + \sqrt{\Delta_{\text{D}}^2 + (J_{\text{S}}-J_{\text{D}})^2}}{\Delta_{\text{D}}}, -1, 1 \right),\nonumber\\
\Psi_3 &= \left( \frac{J_{\text{S}}-J_{\text{D}} - \sqrt{\Delta_{\text{D}}^2 + (J_{\text{S}}-J_{\text{D}})^2}}{\Delta_{\text{D}}}, \frac{J_{\text{S}}-J_{\text{D}} - \sqrt{\Delta_{\text{D}}^2 + (J_{\text{S}}-J_{\text{D}})^2}}{\Delta_{\text{D}}}, 1, 1 \right),\nonumber\\
\Psi_4 &= \left( \frac{J_{\text{S}}-J_{\text{D}} - \sqrt{\Delta_{\text{D}}^2 + (J_{\text{S}}-J_{\text{D}})^2}}{\Delta_{\text{D}}}, -\frac{J_{\text{S}}-J_{\text{D}} - \sqrt{\Delta_{\text{D}}^2 + (J_{\text{S}}-J_{\text{D}})^2}}{\Delta_{\text{D}}}, -1, 1 \right).
\end{align}
At half-filling the bulk band gap closes when $E_2=E_3=0$, which occurs only if $\sqrt{\Delta_{\text{D}}^2 + (J_{\text{S}}-J_{\text{D}})^2} = \Delta_{\text{S}}$. Around this phase transition point, the eigenstate $\Psi_2$ or $\Psi_3$ is occupied, leading to a discontinuous change of the amplitude of the wavefunctions for occupied states.

\subsection{Bulk polarization of the electronic ground state in the low-energy model}
\label{Appendix:LLs2}
In the main text we demonstrated that, due to the symmetry of Eq.~(\ref{Hdirac_AFM}), the $\mathcal{N}>0$ Landau level subspaces do not contribute to the electronic polarization. 
Thus we restrict focus to the Hamiltonian in the $\mathcal{N}=0$ subspace, Eq.~(\ref{Hamiltonian_N=0}), which takes the form of a generalized SSH model.
That is, when we consider an infinite number of layers (\textit{i.e.}~implement periodic boundary conditions) in Eq.~(\ref{supp:Hamiltonian_dirac}), the result takes a tight-binding form \cite{mahon2023reconciling} in the stacking ($\hat{\bm{z}}$) direction.
And when this 3D effective model is projected into the $\mathcal{N}=0$ subspace, the result is a one-dimensional (1D) tight-binding model.
The ground state polarization in our low-energy 3D model has one bulk component, namely $P^{z}$,
which can be calculated using expressions derived in the modern theory of polarization \cite{Vanderbilt1993,Resta2007}.
The $\mathcal{N}=0$ contribution to $P^{z}$ is proportional to the 1D bulk polarization of Eq.~(\ref{Hamiltonian_N=0}) at half-filling, namely
\begin{align}
    P_{\text{SSH}}^{z}=e\sum_{n=1}^{2}\int_{\text{BZ}}\frac{dk_{z}}{2\pi}\xi^{z}_{nn}(k_{z}),
    \label{P0}
\end{align}
where
$\xi^{z}_{nm}(k_{z})\equiv i\braket{u_{n,k_{z}}}{\frac{\partial}{\partial k^{z}}u_{m,k_{z}}}$ are band components of the Berry connection induced by a choice of Bloch energy eigenvectors with cell-periodic parts $\ket{u_{n,k_{z}}}$. 
Inversion symmetry of the model dictates that $P_{\text{SSH}}^{z}\text{ mod } e\in\{0,e/2\}$ \cite{Vanderbilt1993surface}.
A set of analytical eigenvectors of Eq.~(\ref{Hamiltonian_N=0}) can be found and the components differentiated with respect to $k_z$ in an effort to obtain $\xi^{z}_{nn}$.

One issue remains: how do we take the $k_{z}$-derivative of the basis states with respect to which the Hamiltonian matrix (\ref{Hamiltonian_N=0}) is written? 
Typically in phenomenological tight-binding models those basis states are taken to be such that the corresponding orbitals are highly localized in real-space and centered at the ionic positions.
Often the derivatives of these basis states do not affect the results of calculations of topological quantities (e.g., polarization of the usual SSH model or the Chern number of the Haldane model) and one can get away with taking the basis states to be independent of $\boldsymbol{k}$.
However, in the present case accounting for these derivatives becomes important.
For if we assume the basis states to be independent of $k_{z}$ then we find, in general, that $P_{\text{SSH}}^{z}\text{ mod } e\notin\{0,e/2\}$. Thus, that assumption leads to a contradiction and is therefore not valid. 
Instead, we employ approximation Eq.~(\ref{uBarDeriv}) (detailed below) and take $\delta^{z}_{0}=-\delta^{z}_{3}=-d$ and $\delta^{z}_{1}=-\delta^{z}_{2}=0$; we assign the spin-up orbital associated with the bottom (top) surface of the bottom (top) layer the position $z=-d$ ($z=d$) and the spin-up orbital associated with the top (bottom) surface of the bottom (top) layer the position $z=0$ in the unit cell whose center is taken to be the point in the middle of the two layers (which are stacked on top of one another, see Fig.~1 of Ref.~\cite{Lei2020}).
With this, we semianalytically compute Eq.~(\ref{P0}), the results of which are in perfect agreement with Fig.~2 (a) of the main text.
It is notable that in this case the calculation of $P^{z}$ is sensitive to the approximations made to the basis functions, while later in calculations of the topological magnetoelectric coefficient, this is no longer the case.

\subsection{Antiferromagnetic thin films}
\label{Appendix:LLs3}
The phase diagram in Fig.~3 of the main text may be understood in the limit of $\Delta_{\text{D}} \rightarrow 0$. In this case there is no hybridization between electronic states in different van der Waals layers and within each layer the $\mathcal{N}=0$ Hamiltonian generally acts on the surface degree of freedom via a $2\times 2$ matrix:
\begin{equation}
     H_{\text{1L}}^{n=0} = 
 \left(
 \begin{array}{cccc}
       m_t & \Delta_{\text{S}}  \\
       \Delta_{\text{S}} & m_b \\
 \end{array}
     \right) = \frac{m_t + m_b}{2}\left(
 \begin{array}{cccc}
       1 & 0  \\
       0 & 1 \\
 \end{array}
     \right) + 
     \left(
 \begin{array}{cccc}
       (m_t -m_b)/2 & \Delta_{\text{S}}  \\
       \Delta_{\text{S}} & -(m_t-m_b)/2 \\
 \end{array}
     \right),
\end{equation}
which may be written as $H_{\text{1L}} = \bar{m}\tau_0 + \tilde{m} \tau_z + \Delta_{\text{S}} \tau_x$ if we define $\bar{m} \equiv (m_t + m_b)/2$ and $\tilde{m} \equiv (m_t - m_b)/2$. Here $\tau_{i}$ denote the Pauli matrices. On the outside surfaces $m_{t/b} = \pm J_{\text{S}}$ whereas on the interior surfaces $m_{t/b} = \pm (1-\delta)J_{\text{S}}$ with $\delta \equiv J_{\text{D}}/J_{\text{S}}$. 
In the limit of $\delta \rightarrow 0$, the Hamiltonian in the $\mathcal{N}=0$ subspace have for a two layer thin film is
\begin{equation}
     H_{\text{2L}}^{n=0} = 
 \left(
 \begin{array}{cccc}
       J_{\text{S}} & \Delta_{\text{S}} & 0 & 0 \\
       \Delta_{\text{S}} & J_{\text{S}} & 0 & 0 \\
       0 & 0 & -J_{\text{S}} & \Delta_{\text{S}} \\
       0 & 0 & \Delta_{\text{S}} & -J_{\text{S}} \\
 \end{array}
     \right).
\end{equation}
The eigenvalues are $\pm J_{\text{S}} \pm \Delta_{\text{S}}$ and thus have a phase transition at $J_{\text{S}} = \Delta_{\text{S}}$. 
If $J_{\text{S}} < \Delta_{\text{S}}$, the two states with negative energies are $E_- = \pm J_{\text{S}} - \Delta_{\text{S}}$ which belong different layers, else if $J_{\text{S}} > \Delta_{\text{S}}$, the two states with negative energies are $E_- = -J_{\text{S}} \pm \Delta_{\text{S}}$ which belong to the same layer.

In the case of $\delta \ne 0$, the layers in $N$-layer thin films may be grouped into the outer layers and interior layers. As an illustration, the following Hamiltonian describes a four-layer thin film:
\begin{equation}
  H^{n=0}_{\text{4L}} = \left(
      \begin{array}{cccccccc}
        J_{\text{S}} &  \Delta_{\text{S}} & 0  & 0 & 0  & 0 & 0  & 0 \\
        \Delta_{\text{S}} & J_{\text{S}}-J_{\text{D}} & 0 & 0 & 0  & 0 & 0  & 0 \\
        0 & 0 & -J_{\text{S}}+J_{\text{D}} & \Delta_{\text{S}} & 0  & 0 & 0  & 0  \\
        0 & 0 & \Delta_{\text{S}} & -J_{\text{S}}+J_{\text{D}} & 0  & 0 & 0  & 0 \\
        0 & 0  & 0 & 0  & J_{\text{S}}-J_{\text{D}} & \Delta_{\text{S}}  & 0 & 0  \\
        0 & 0  & 0 & 0  & \Delta_{\text{S}} & J_{\text{S}}-J_{\text{D}}  & 0 & 0  \\
        0 & 0  & 0 & 0  & 0 & 0  & -J_{\text{S}}+J_{\text{D}} & \Delta_{\text{S}}  \\
        0 & 0  & 0 & 0  & 0 & 0  & \Delta_{\text{S}} & -J_{\text{S}}  \\
      \end{array}
      \right).
      \label{Hamiltonian_N=0_film}
\end{equation}
Generalized to $N$-layer thin films, the energy eigenvalues and (nonnormalized) eigenstates can be solved as
\begin{align}\label{eigensystem_NL}
& E_1 = (1-\delta/2) J_{\text{S}} + \sqrt{\Delta_{\text{S}}^2 + (J_{\text{S}}\delta/2)^2}; && \Psi_1 = \big( \phi_{+}, 1, 0, 0, \cdots ,0, 0, 0, 0 \big), \nonumber\\
& E_2 = (1-\delta/2) J_{\text{S}} - \sqrt{\Delta_{\text{S}}^2 + (J_{\text{S}}\delta/2)^2}; && \Psi_2 = \big( \phi_{-}, 1, 0, 0, \cdots, 0, 0, 0, 0 \big), \nonumber\\
& E_3 = -(1-\delta) J_{\text{S}} + \Delta_{\text{S}}; && \Psi_3 = \big( 0, 0, 1, 1, \cdots, 0, 0, 0, 0 \big), \nonumber\\
& E_4 = -(1-\delta) J_{\text{S}} - \Delta_{\text{S}}; && \Psi_4 = \big( 0, 0, -1, 1, \cdots, 0, 0, 0, 0 \big), \nonumber\\
& \cdots \cdots &&   \cdots \cdots \nonumber\\
& E_{2N-3} = (1-\delta) J_{\text{S}} + \Delta_{\text{S}}; && \Psi_{2N-3} = \big( 0, 0, 0, 0, \cdots, 1, 1, 0, 0 \big), \nonumber\\
& E_{2N-2} = (1-\delta) J_{\text{S}} - \Delta_{\text{S}}; && \Psi_{2N-2} = \big( 0, 0, 0, 0, \cdots, -1, 1, 0, 0 \big), \nonumber\\
& E_{2N-1} = -(1-\delta/2) J_{\text{S}} + \sqrt{\Delta_{\text{S}}^2 + (J_{\text{S}}\delta/2)^2}; && \Psi_{2N-1} = \big( 0, 0, 0, 0, \cdots, 0, 0, \phi_{+}, 1 \big), \nonumber\\
& E_{2N} = -(1-\delta/2) J_{\text{S}} - \sqrt{\Delta_{\text{S}}^2 + (J_{\text{S}}\delta/2)^2}; && \Psi_{2N} = \big( 0, 0, 0, 0, \cdots, 0, 0, \phi_{-}, 1 \big).
\end{align}
Here $\phi_{\pm} \equiv J_{\text{S}}\delta/2\Delta_{\text{S}} \pm \sqrt{1 + (J_{\text{S}}\delta/2\Delta_{\text{S}})^2}$, the normalization factor of the wavefunctions are 
$\mathcal{A}_{\pm} = 1/\sqrt{1 + |\phi_{\pm}|^2}$ and $1/\sqrt{2}$. Note that in the limit of $\delta \to 0$, $\phi_{\pm} \to \pm 1$ and $\mathcal{A}_{\pm} \to 1/\sqrt{2}$. When setting the center of the thin films as the origin ($z=0$) and ascribing to the top and bottom surface of each layer the position $z_j = jd_s + (j-1/2)d_v$ and $z_j = (j-1)d_s + (j-1/2)d_v$ for $z>0$, whereas $z_j = -jd_s - (j-1/2)d_v$ and $z_j = -(j-1)d_s - (j-1/2)d_v$ for $z<0$, with $d_s$ the thickness of each layer (distance between the top and bottom surface for each layer) and $d_v$ the distance between each layer. Here $j = 1,2,\ldots,N/2$ where $N$ is the total number of layers of the thin films.

As shown in the main text, the magnetoelectric coefficient may be written as
\begin{equation}
    \alpha_{\text{me}} = \frac{e^2}{N(d_s + d_v)hc} \sum_{j,E<E_F} |C_{E}(z_j)|^2 z_j.
\end{equation}
Here $E_F$ is the Fermi energy which is $0$ in our model. In the following we will discuss the magnetoelectric coefficient for parameters in different regions:
\begin{enumerate}
\item $J_{\text{S}} < \Delta_{\text{S}}$ \\
For $J_{\text{S}} < \Delta_{\text{S}}$, $E_{2n} < 0$ in Eq.~(\ref{eigensystem_NL}) and thus the contribution of interior layers is 0 since $|C_{E}(z_j)|^2 = C_{E}(-z_j)|^2$. 
\begin{equation}
    \alpha_{\text{me}}|_{J_{\text{S}} < \Delta_{\text{S}}} = \frac{e^2}{Nhc} \frac{d_s}{d_s + d_v}\mathcal{A}_-^2 (|\phi_{-}|^2-1),
\end{equation} 
which is always equal to $0$ when $N \to \infty$ or $\delta \to 0$.

\item $J_{\text{S}} > \Delta_{\text{S}}$ \\
For the case of $J_{\text{S}} > \Delta_{\text{S}}$, as we show in the main text there are two phase transition points when varying the value of $\delta$, it is easy to see in the band gap which closes two times.
  \begin{enumerate}
      \item $\delta < 1- \Delta_{\text{S}}/J_{\text{S}}$ \\
       In this case $(1-\delta/2) J_{\text{S}} - \sqrt{\Delta_{\text{S}}^2 + (J_{\text{S}}\delta/2)^2} >0$ and $(1-\delta) J_{\text{S}} - \Delta_{\text{S}} > 0$, layers with even number is occupied, and thus the contribution from the interior layers is $\alpha_{\text{me}}^j = e^2(2j-1)/Nhc$. For outer layers, the contribution is:
       \begin{equation}
        \alpha_{\text{me}}^{\text{out}} = 2\frac{e^2}{hc} \frac{N-1}{2N}
         + \frac{d_s}{2N(d_s + d_v)} (\mathcal{A}_{+}^2 + \mathcal{A}_{-}^2 - \mathcal{A}_{+}^2 \phi_{+}^2 - \mathcal{A}_{-}^2 \phi_{-}^2),
       \end{equation}
       where we used the fact that $\mathcal{A}_{+}^2 + \mathcal{A}_{-}^2 + \mathcal{A}_{+}^2 \phi_{+}^2 + \mathcal{A}_{-}^2 \phi_{-}^2 = 2$ since $\mathcal{A}_{\pm} = 1/\sqrt{1+\phi_{\pm}^2}$. The total magnetoelectric coefficient is thus:
       \begin{equation}
           \alpha_{\text{me}}|_{J_{\text{S}}>\Delta_{\text{S}},\delta < 1- \Delta_{\text{S}}/J_{\text{S}}} = \alpha_{\text{me}}^{\text{out}} + \sum_{j=1}^{N/2-1} (-1)^{N/2+j}\alpha_{\text{me}}^j. 
       \end{equation}
       Here the factor $(-1)^{N/2+j}$ is introduced so that the sign of $\alpha_{\text{me}}$ keeps the same different thickness of thin films. In the limit of $\delta \to 0$, the magnetoelectric coefficient becomes:
       \begin{equation}
           \alpha_{\text{me}}|_{J_{\text{S}}>\Delta_{\text{S}},\delta \to 0} = \frac{e^2}{2hc}\sum_{j=1}^{N/2} (-1)^{N/2+j}\frac{2j-1}{N/2}. 
       \end{equation}
       
      \item $1- \Delta_{\text{S}}/J_{\text{S}} < \delta < 1- (\Delta_{\text{S}}/J_{\text{S}})^2$ \\
      In this case, $(1-\delta/2) J_{\text{S}} - \sqrt{\Delta_{\text{S}}^2 + (J_{\text{S}}\delta/2)^2} >0$ and $(1-\delta) J_{\text{S}} - \Delta_{\text{S}} < 0$, the interior layer contribution is zero as $|C_{E}(z_j)|^2 = C_{E}(-z_j)|^2$. The contribution to the magnetoelectric coefficient is thus only from the outer layer:
      \begin{equation}
        \alpha_{\text{me}} = \alpha_{\text{me}}^{\text{out}} = 2\frac{e^2}{hc} \frac{N-1}{2N}
         + \frac{d_s}{2N(d_s + d_v)} (\mathcal{A}_{+}^2 + \mathcal{A}_{-}^2 - \mathcal{A}_{+}^2 \phi_{+}^2 - \mathcal{A}_{-}^2 \phi_{-}^2),
       \end{equation}
       which is $e^2/hc$ in the limit of $N \to \infty$.
      \item $\delta > (\Delta_{\text{S}}/J_{\text{S}})^2$ \\
      In this case, $(1-\delta/2) J_{\text{S}} - \sqrt{\Delta_{\text{S}}^2 + (J_{\text{S}}\delta/2)^2} <0$ and $(1-\delta) J_{\text{S}} - \Delta_{\text{S}} < 0$, the occupation of the states is exactly the same as the case of $J_{\text{S}} < \Delta_{\text{S}}$, and the magnetoelectric response goes to 0 in the thick limit of thin films.
      
  \end{enumerate}
\end{enumerate}

\begin{figure}[t!]
  \centering
  \includegraphics[width=0.9 \linewidth ]{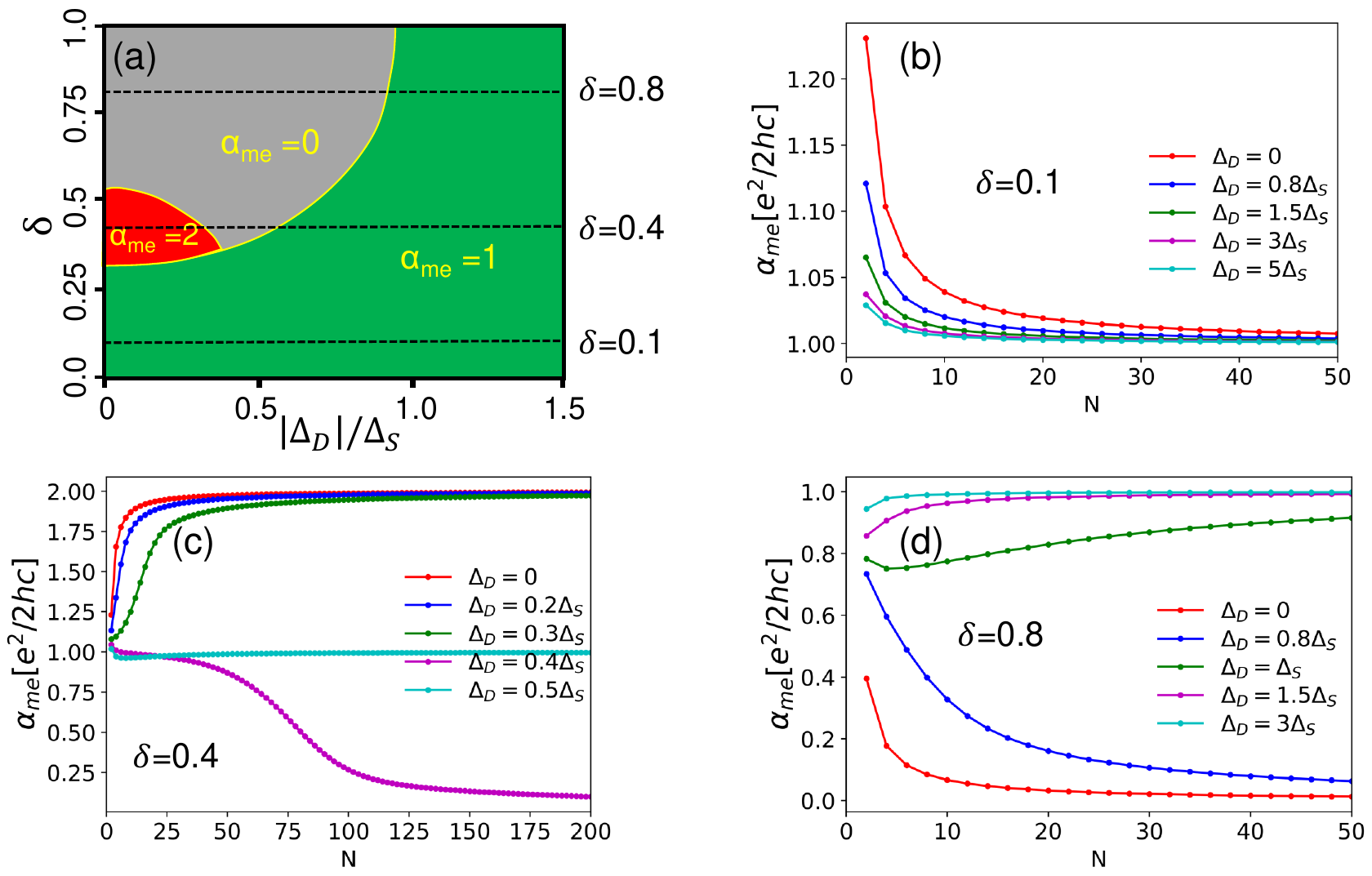}
  \caption{
  (a) Thin film phase diagram for $J_{\text{S}} =1.5 \Delta_{\text{S}}$ as discussed in the main text. Panels (b)-(d) are the plots of the magnetoelectric coefficient versus the thickness of thin films for various $\delta$ and $\Delta_{\text{D}}$, with the values of $\delta$ labeled with black dashed lines in panel (a). 
  }\label{fig:ME_vs_thickness}
\end{figure}

For general values of $\delta$ and $\Delta_{\text{D}}$, the magnetoelectric coefficients calculated from numerical methods are shown in Figs.~\ref{fig:ME_vs_thickness}(b)-\ref{fig:ME_vs_thickness}(d), and the corresponding phase is shown in Fig.~\ref{fig:ME_vs_thickness}(a).

\section{Bulk calculations for three-dimensional MBX-type crystalline insulators}
\label{Appendix:3DTightBinding}
\subsection{Tight-binding model}
\label{Appendix:3DTightBinding1}
We employ a previously developed \cite{mahon2023reconciling} square lattice regularization for the effective low-energy coupled Dirac cone model that was introduced (Eq.~(2)) in the main text.
The family of materials we consider are layered compounds (with stacking direction taken to be $\bm{z}$ and interlayer spacing $d$), and the tight-binding model we use was obtained by developing a 2D square lattice regularization of the $\boldsymbol{k}\cdot\boldsymbol{p}$ model of an isolated layer then coupling the states associated with the nearest surfaces of adjacent layers.
In the nonmagnetic case, the model is periodic under translations by $d\bm{z}$ (\textit{i.e.}, one layer per unit cell), while in the AFM case the model is periodic under translations by $2d\bm{z}$ (\textit{i.e.}, two layers per unit cell).
The 3D square lattice regularized model (with Bravais lattice $\Gamma=\mathbb{Z}\times\mathbb{Z}\times 2d\mathbb{Z}$ and whose dual lattice is $\Gamma^{*}=2\pi\mathbb{Z}\times2\pi\mathbb{Z}\times \frac{2\pi}{2d}\mathbb{Z}$) can be written in the general form
\begin{equation}
    \hat{H}^{(\text{3D})}_{\text{reg}}=\int_{\text{BZ}}\frac{d^3k}{(2\pi)^3}\nonumber \hat{c}^{\dagger}(\boldsymbol{k}) \mathcal{H}^{(\text{3D})}_{\text{reg}}(\boldsymbol{k})\hat{c}(\boldsymbol{k}),
\end{equation}
where $\mathcal{H}^{(\text{3D})}_{\text{reg}}(\boldsymbol{k})$ is a smooth, Hermitian-matrix-valued map in $\mathbb{R}^{3}$ that is $\Gamma^{*}$-periodic (\textit
{i.e.}, $\forall\boldsymbol{G}\in\Gamma^{*}:\mathcal{H}^{(\text{3D})}_{\text{reg}}(\boldsymbol{k})=\mathcal{H}^{(\text{3D})}_{\text{reg}}(\boldsymbol{k}+\boldsymbol{G})$ is satisfied), BZ denotes a 3D Brillouin zone (BZ) of $\Gamma^{*}$, and
\begin{align}
\hat{c}(\boldsymbol{k})&\equiv \left(\hat{c}_{(0,\uparrow),\boldsymbol{k}},\hat{c}_{(0,\downarrow),\boldsymbol{k}},\hat{c}_{(1,\uparrow),\boldsymbol{k}},\hat{c}_{(1,\downarrow),\boldsymbol{k}},\hat{c}_{(2,\uparrow),\boldsymbol{k}},\hat{c}_{(2,\downarrow),\boldsymbol{k}},\hat{c}_{(3,\uparrow),\boldsymbol{k}},\hat{c}_{(3,\downarrow),\boldsymbol{k}}\right)^{\text{T}}, \nonumber\\ 
\hat{c}^{\dagger}(\boldsymbol{k})&\equiv \left(\hat{c}^{\dagger}_{(0,\uparrow),\boldsymbol{k}},\hat{c}^{\dagger}_{(0,\downarrow),\boldsymbol{k}},\hat{c}^{\dagger}_{(1,\uparrow),\boldsymbol{k}},\hat{c}^{\dagger}_{(1,\downarrow),\boldsymbol{k}},\hat{c}^{\dagger}_{(2,\uparrow),\boldsymbol{k}},\hat{c}^{\dagger}_{(2,\downarrow),\boldsymbol{k}},\hat{c}^{\dagger}_{(3,\uparrow),\boldsymbol{k}},\hat{c}^{\dagger}_{(3,\downarrow),\boldsymbol{k}}\right).
\end{align}
The operators $\hat{c}_{(\alpha,\sigma),\boldsymbol{k}}$, $\hat{c}^{\dagger}_{(\alpha,\sigma),\boldsymbol{k}}$ correspond to Block-type vectors $\ket{\bar{\psi}_{(\alpha,\sigma),\boldsymbol{k}}}\equiv\hat{c}^{\dagger}_{(\alpha,\sigma),\boldsymbol{k}}\ket{\text{vac}}$ (\textit{i.e.}~they satisfy $\ket{\bar{\psi}_{(\alpha,\sigma),\boldsymbol{k}+\boldsymbol{G}}}=\ket{\bar{\psi}_{(\alpha,\sigma),\boldsymbol{k}}}$ for any $\bm{G}\in\Gamma^{*}$) that are smooth over BZ, are mutually orthogonal, and map to the exponentially localized Wannier functions (WFs) $\ket{W_{(\alpha,\sigma),\boldsymbol{R}}}$ \cite{Marzari2012} with respect to which the lattice regularized tight-binding model is written.
We use even (odd) values of $\alpha$ to label WFs that are associated with the bottom (top) surface \cite{zhang2009topological,Wang2019} of each layer.
If there are $L$ layers in a unit cell then the WFs $\ket{W_{(\alpha,\sigma),\boldsymbol{R}}}$ associated with layer $l\in\{1,\ldots,L\}$ (enumerated from bottom layer to top layer) are labeled by $\alpha=2(l-1)$ and $2l-1$.
We denote the corresponding cell-periodic parts of the Bloch-type vectors $\ket{\bar{\psi}_{(\alpha,\sigma),\boldsymbol{k}}}$ by $\ket{\bar{u}_{(\alpha,\sigma),\boldsymbol{k}}}$.

As previously described \cite{mahon2023reconciling}, the $\ket{W_{(\alpha,\sigma),\boldsymbol{R}}}$ can be assumed to be atomic-like (this is almost always done without mention in tight-binding calculations), a characteristic property of which is $\partial_{a}\ket{\bar{u}_{(\alpha,\sigma),\boldsymbol{k}}}\equiv 0$ where $\partial_{a}\equiv \partial/\partial k^{a}$ and $a=1$, $2$, or $3$ ($\equiv x$, $y$, or $z$).
A more physical (although still highly simplified) approximation is reached by assuming that each WF $W_{(i,\alpha),\boldsymbol{R}}(\boldsymbol{r})$ is extremely localized about a point $\boldsymbol{\delta}_{\alpha}$ in the unit cell at $\boldsymbol{R}$ with which it is associated, from which it follows \cite{Muniz}
\begin{gather}
\frac{\partial}{\partial k^a}\bar{u}_{(\alpha,\sigma),\boldsymbol{k}}(\boldsymbol{r})=\sqrt{\frac{\Omega_{uc}}{(2\pi)^3}}\sum_{\boldsymbol{R}}\frac{\partial e^{-i\boldsymbol{k}\cdot(\boldsymbol{r}-\boldsymbol{R})}}{\partial k^a}W_{(\alpha,\sigma),\boldsymbol{R}}(\boldsymbol{r})\approx\sqrt{\frac{\Omega_{uc}}{(2\pi)^3}}\sum_{\boldsymbol{R}}\frac{\partial e^{-i\boldsymbol{k}\cdot\boldsymbol{\delta}_{\alpha}}}{\partial k^a}W_{(\alpha,\sigma),\boldsymbol{R}}(\boldsymbol{r})=-i\delta_{\alpha}^a\bar{u}_{(\alpha,\sigma),\boldsymbol{k}}(\boldsymbol{r}).
\label{uBarDeriv}
\end{gather}
In the nonmagnetic case, there is one layer per unit cell and if the layer thickness is neglected then one can choose $\bm{\delta}_{\alpha}=\bm{0}$, which is equivalent to the assumption of atomic-like WFs.
However, in the AFM case there are two layers per unit cell, thus there is always more than one position within it about which WFs are localized. 
In particular, even if the WFs within each layer are taken to be localized about the same point, $\boldsymbol{\delta}_{0}=\boldsymbol{\delta}_{1}$ and $\boldsymbol{\delta}_{2}=\boldsymbol{\delta}_{3}$, we have $\boldsymbol{\delta}_{2}-\boldsymbol{\delta}_{0}=d\boldsymbol{z}$. 
In this approximation, the most natural choice is to take $\boldsymbol{\delta}_{2}=\frac{d}{2}\boldsymbol{z}$ and $\boldsymbol{\delta}_{0}=-\frac{d}{2}\boldsymbol{z}$.

Under our lattice regularization scheme, the general AFM Hamiltonian matrix is
\small
\begin{align}
&\mathcal{H}^{(\text{3D})}_{\text{reg}}(\boldsymbol{k}) = \nonumber\\
&\left(\begin{array}{cccccccc}
    \mu & iA(\text{s}k_x-i\text{s}k_y) & \Delta_{\text{S}}(k_x,k_y) & 0 & 0 & 0 & e^{-2idk_{z}}\Delta_{\text{D}} & 0	\\
    - iA(\text{s}k_x+i\text{s}k_y) & -\mu & 0 & \Delta_{\text{S}}(k_x,k_y) & 0 & 0 & 0 & e^{-2idk_{z}}\Delta_{\text{D}}	\\
    \Delta_{\text{S}}(k_x,k_y) & 0 & \mu & - iA(\text{s}k_x-i\text{s}k_y) & \Delta_{\text{D}} & 0 & 0 & 0	\\
    0 & \Delta_{\text{S}}(k_x,k_y) & iA(\text{s}k_x+i\text{s}k_y) & -\mu & 0 & \Delta_{\text{D}} & 0 & 0 \\
    0 & 0 & \Delta_{\text{D}} & 0 & -\mu & iA(\text{s}k_x-i\text{s}k_y) & \Delta_{\text{S}}(k_x,k_y) & 0	\\
    0 & 0 & 0 & \Delta_{\text{D}} & - iA(\text{s}k_x+i\text{s}k_y) & \mu & 0 & \Delta_{\text{S}}(k_x,k_y)	\\
    e^{2idk_{z}}\Delta_{\text{D}} & 0 & 0 & 0 & \Delta_{\text{S}}(k_x,k_y) & 0 & -\mu & - iA(\text{s}k_x-i\text{s}k_y)	\\
    0 & e^{2idk_{z}}\Delta_{\text{D}} & 0 & 0 & 0 & \Delta_{\text{S}}(k_x,k_y) & iA(\text{s}k_x+i\text{s}k_y) & \mu
\end{array}\right).
\label{H_AFM}
\end{align}
\normalsize
Here $\Delta_{\text{S}}(k_x,k_y)\equiv\Delta_{\text{S}}-2B(2-\text{c}k_{x}-\text{c}k_{y})$, where $\text{c}k_x\equiv\cos(k_{x})$, $\text{s}k_x\equiv\sin(k_{x})$, etc. This Hamiltonian is valid only when $\mu\neq0$, since otherwise the material is nonmagnetic and the crystalline unit cell is misidentified \cite{mahon2023reconciling}. The eigenvalues of Eq.~(\ref{H_AFM}) are 
\begin{align}
E_{1,2}(\boldsymbol{k})&=-	E_{7,8}(\boldsymbol{k})=-\sqrt{\frac{1}{2}A^2(2-\text{c}2k_{x}-\text{c}2k_{y})+\Delta_{\text{S}}(k_{x},k_{y})^2+\Delta_{\text{D}}^2+\mu^2+2\sqrt{\Delta_{\text{S}}(k_{x},k_{y})^2(\Delta_{\text{D}}^2\cos^2(dk_z)+\mu^2)}},\nonumber\\ 
E_{3,4}(\boldsymbol{k})&=-E_{5,6}(\boldsymbol{k})=-\sqrt{\frac{1}{2}A^2(2-\text{c}2k_{x}-\text{c}2k_{y})+\Delta_{\text{S}}(k_{x},k_{y})^2+\Delta_{\text{D}}^2+\mu^2-2\sqrt{\Delta_{\text{S}}(k_{x},k_{y})^2(\Delta_{\text{D}}^2\cos^2(dk_z)+\mu^2)}}.
\label{eigenvaluesAFM}
\end{align}
We will consider the ground state of the material to be that for which the energy bands are half-filled. We restrict our focus to band insulators, which implies $\Delta_{\text{S}}^2+\Delta_{\text{D}}^2+\mu^2-2\sqrt{\Delta_{\text{S}}^2(\Delta_{\text{D}}^2+\mu^2)} > 0$ (\textit{i.e.}, $|\Delta_{\text{S}}|-\sqrt{\Delta_{\text{D}}^2+\mu^2}\neq 0$).
Moreover, the family of MBX materials of interest are such that the band gap is smallest at $\bm{k}=(0,0,0)$, which is always the case when $B/\Delta_{\text{S}}<0$.
The double degeneracy of the energy bands at each $\boldsymbol{k}\in\text{BZ}$ follows from the combination of an inversion and a (fermionic) time-reversal symmetry (see Appendix C of Ref.~\cite{mahon2023reconciling}). A consequence is that the corresponding eigenvectors are highly nonunique. We wish to identify a set of orthogonal energy eigenvectors that are smooth and orthogonal over BZ such that they constitute a smooth global (and periodic) gauge choice of the bundle of occupied Bloch states over the Brillouin zone torus and of the total Bloch bundle more generally. One such choice, written in the basis $\big(\ket{\bar{\psi}_{(\alpha,\uparrow),\boldsymbol{k}}},\,\ket{\bar{\psi}_{(\alpha,\downarrow),\boldsymbol{k}}}\big)_{\alpha\in\{0,1,2,3\}}$, is
\small
\begin{subequations}
\begin{align}
\Psi_{1,\boldsymbol{k}}&=\Bigg(-\frac{(1 + e^{2 i d k_{z}}) \Delta_{\text{S}}(k_{x},k_{y}) + 
  \sqrt{2 (\Delta_{\text{D}}^2 + 
      2 \mu^2 + \Delta_{\text{D}}^2 \text{c}(2 d k_{z}))}}{
 (-1 + e^{2 i d k_{z}}) \mu - (1 + e^{2 i d k_{z}})E_{1}(\boldsymbol{k})}, \, 0, \, 1, \, \frac{
 (1 + e^{2 i d k_{z}}) A (-i \text{s}k_{x} + \text{s}k_{y})}{
  (-1 + e^{2 i d k_{z}}) \mu - (1 + e^{2 i d k_{z}})E_{1}(\boldsymbol{k})},\nonumber\\
&\qquad-\frac{(1 + e^{2 i d k_{z}}) \Delta_{\text{D}}^2 + 
     e^{2 i d k_{z}} \Big(2 \mu(\mu-E_{1}(\boldsymbol{k})) + 
        \sqrt{\Delta_{\text{S}}(k_{x},k_{y})^2 (2 (\Delta_{\text{D}}^2 + 
             2 \mu^2 + \Delta_{\text{D}}^2 \text{c}(2 d k_{z})))}\Big)}{\Delta_{\text{D}} ( (-1 + e^{2 i d k_{z}}) \mu - (1 + e^{2 i d k_{z}})E_{1}(\boldsymbol{k}))}, \nonumber\\
&\qquad \frac{
 2e^{2 i d k_{z}}
   A \mu (-i \text{s}k_{x} + 
    \text{s}k_{y})}{\Delta_{\text{D}} \big( (-1 + e^{2 i d k_{z}}) \mu - (1 + e^{2 i d k_{z}}) E_{1}(\boldsymbol{k}) \big)}, \, 
\frac{2e^{2 i d k_{z}} \big(\Delta_{\text{S}}(k_{x},k_{y}) \mu + 
    \sqrt{\mu^2 + \Delta_{\text{D}}^2 \text{c}(d k_{z})^2} (\mu - 
       E_{1}(\boldsymbol{k}))\big)}{\Delta_{\text{D}} \big( (-1 + e^{2 i d k_{z}}) \mu - (1 + e^{2 i d k_{z}}) E_{1}(\boldsymbol{k})\big)},\nonumber\\
&\qquad\frac{2 e^{2 i d k_{z}}
   A \sqrt{\mu^2 + \Delta_{\text{D}}^2 \text{c}(dk_{z})^2} (-i \text{s}k_{x} + 
    \text{s}k_{y})}{\Delta_{\text{D}} \big((-1 + e^{2 i d k_{z}}) \mu - (1 + e^{2 i d k_{z}})E_{1}(\boldsymbol{k})\big)}\Bigg), \nonumber\\
\Psi_{7,\boldsymbol{k}}&=\Bigg(-\frac{(1 + e^{2 i d k_{z}}) \Delta_{\text{S}}(k_{x},k_{y}) + 
  \sqrt{2 (\Delta_{\text{D}}^2 + 
      2 \mu^2 + \Delta_{\text{D}}^2 \text{c}(2 d k_{z}))}}{
 (-1 + e^{2 i d k_{z}}) \mu - (1 + e^{2 i d k_{z}})E_{7}(\boldsymbol{k})}, \, 0, \, 1, \, \frac{
 (1 + e^{2 i d k_{z}}) A (-i \text{s}k_{x} + \text{s}k_{y})}{
  (-1 + e^{2 i d k_{z}}) \mu - (1 + e^{2 i d k_{z}})E_{7}(\boldsymbol{k})},\nonumber\\
&\qquad-\frac{(1 + e^{2 i d k_{z}}) \Delta_{\text{D}}^2 + 
     e^{2 i d k_{z}} \Big(2 \mu(\mu-E_{7}(\boldsymbol{k})) + 
        \sqrt{\Delta_{\text{S}}(k_{x},k_{y})^2 (2 (\Delta_{\text{D}}^2 + 
             2 \mu^2 + \Delta_{\text{D}}^2 \text{c}(2 d k_{z})))}\Big)}{\Delta_{\text{D}} ( (-1 + e^{2 i d k_{z}}) \mu - (1 + e^{2 i d k_{z}})E_{7}(\boldsymbol{k}))}, \nonumber\\
&\qquad \frac{
 2e^{2 i d k_{z}}
   A \mu (-i \text{s}k_{x} + 
    \text{s}k_{y})}{\Delta_{\text{D}} \big( (-1 + e^{2 i d k_{z}}) \mu - (1 + e^{2 i d k_{z}}) E_{7}(\boldsymbol{k}) \big)}, \, 
\frac{2e^{2 i d k_{z}} \big(\Delta_{\text{S}}(k_{x},k_{y}) \mu + 
    \sqrt{\mu^2 + \Delta_{\text{D}}^2 \text{c}(d k_{z})^2} (\mu - 
       E_{7}(\boldsymbol{k}))\big)}{\Delta_{\text{D}} \big( (-1 + e^{2 i d k_{z}}) \mu - (1 + e^{2 i d k_{z}}) E_{7}(\boldsymbol{k})\big)},\nonumber\\
&\qquad\frac{2 e^{2 i d k_{z}}
   A \sqrt{\mu^2 + \Delta_{\text{D}}^2 \text{c}(dk_{z})^2} (-i \text{s}k_{x} + 
    \text{s}k_{y})}{\Delta_{\text{D}} \big((-1 + e^{2 i d k_{z}}) \mu - (1 + e^{2 i d k_{z}})E_{7}(\boldsymbol{k})\big)}\Bigg),
\end{align}
\begin{align}
\Psi_{3,\boldsymbol{k}}&=\Bigg(\frac{-(1 + e^{2 i d k_{z}}) \Delta_{\text{S}}(k_{x},k_{y}) + 
  \sqrt{2 (\Delta_{\text{D}}^2 + 
      2 \mu^2 + \Delta_{\text{D}}^2 \text{c}(2 d k_{z}))}}{
 (-1 + e^{2 i d k_{z}}) \mu - (1 + e^{2 i d k_{z}})E_{3}(\boldsymbol{k})}, \, 0, \, 1, \, \frac{
 (1 + e^{2 i d k_{z}}) A (-i \text{s}k_{x} + \text{s}k_{y})}{
  (-1 + e^{2 i d k_{z}}) \mu - (1 + e^{2 i d k_{z}})E_{3}(\boldsymbol{k})},\nonumber\\
&\qquad-\frac{(1 + e^{2 i d k_{z}}) \Delta_{\text{D}}^2 + 
     e^{2 i d k_{z}} \Big(2 \mu(\mu-E_{3}(\boldsymbol{k})) - 
        \sqrt{\Delta_{\text{S}}(k_{x},k_{y})^2 (2 (\Delta_{\text{D}}^2 + 
             2 \mu^2 + \Delta_{\text{D}}^2 \text{c}(2 d k_{z})))}\Big)}{\Delta_{\text{D}} ( (-1 + e^{2 i d k_{z}}) \mu - (1 + e^{2 i d k_{z}})E_{3}(\boldsymbol{k}))}, \nonumber\\
&\qquad \frac{
 2e^{2 i d k_{z}}
   A \mu (-i \text{s}k_{x} + 
    \text{s}k_{y})}{\Delta_{\text{D}} \big( (-1 + e^{2 i d k_{z}}) \mu - (1 + e^{2 i d k_{z}}) E_{3}(\boldsymbol{k}) \big)}, \, 
\frac{2e^{2 i d k_{z}} \big(\Delta_{\text{S}}(k_{x},k_{y}) \mu - 
    \sqrt{\mu^2 + \Delta_{\text{D}}^2 \text{c}(d k_{z})^2} (\mu - 
       E_{3}(\boldsymbol{k}))\big)}{\Delta_{\text{D}} \big( (-1 + e^{2 i d k_{z}}) \mu - (1 + e^{2 i d k_{z}}) E_{3}(\boldsymbol{k})\big)},\nonumber\\
&\qquad -\frac{2 e^{2 i d k_{z}}
   A \sqrt{\mu^2 + \Delta_{\text{D}}^2 \text{c}(dk_{z})^2} (-i \text{s}k_{x} + 
    \text{s}k_{y})}{\Delta_{\text{D}} \big((-1 + e^{2 i d k_{z}}) \mu - (1 + e^{2 i d k_{z}})E_{3}(\boldsymbol{k})\big)}\Bigg),\nonumber\\
\Psi_{5,\boldsymbol{k}}&=\Bigg(\frac{-(1 + e^{2 i d k_{z}}) \Delta_{\text{S}}(k_{x},k_{y}) + 
  \sqrt{2 (\Delta_{\text{D}}^2 + 
      2 \mu^2 + \Delta_{\text{D}}^2 \text{c}(2 d k_{z}))}}{
 (-1 + e^{2 i d k_{z}}) \mu - (1 + e^{2 i d k_{z}})E_{5}(\boldsymbol{k})}, \, 0, \, 1, \, \frac{
 (1 + e^{2 i d k_{z}}) A (-i \text{s}k_{x} + \text{s}k_{y})}{
  (-1 + e^{2 i d k_{z}}) \mu - (1 + e^{2 i d k_{z}})E_{5}(\boldsymbol{k})},\nonumber\\
&\qquad-\frac{(1 + e^{2 i d k_{z}}) \Delta_{\text{D}}^2 + 
     e^{2 i d k_{z}} \Big(2 \mu(\mu-E_{5}(\boldsymbol{k})) - 
        \sqrt{\Delta_{\text{S}}(k_{x},k_{y})^2 (2 (\Delta_{\text{D}}^2 + 
             2 \mu^2 + \Delta_{\text{D}}^2 \text{c}(2 d k_{z})))}\Big)}{\Delta_{\text{D}} ( (-1 + e^{2 i d k_{z}}) \mu - (1 + e^{2 i d k_{z}})E_{5}(\boldsymbol{k}))}, \nonumber\\
&\qquad \frac{
 2e^{2 i d k_{z}}
   A \mu (-i \text{s}k_{x} + 
    \text{s}k_{y})}{\Delta_{\text{D}} \big( (-1 + e^{2 i d k_{z}}) \mu - (1 + e^{2 i d k_{z}}) E_{5}(\boldsymbol{k}) \big)}, \, 
\frac{2e^{2 i d k_{z}} \big(\Delta_{\text{S}}(k_{x},k_{y}) \mu - 
    \sqrt{\mu^2 + \Delta_{\text{D}}^2 \text{c}(d k_{z})^2} (\mu - 
       E_{5}(\boldsymbol{k}))\big)}{\Delta_{\text{D}} \big( (-1 + e^{2 i d k_{z}}) \mu - (1 + e^{2 i d k_{z}}) E_{5}(\boldsymbol{k})\big)},\nonumber\\
&\qquad -\frac{2 e^{2 i d k_{z}}
   A \sqrt{\mu^2 + \Delta_{\text{D}}^2 \text{c}(dk_{z})^2} (-i \text{s}k_{x} + 
    \text{s}k_{y})}{\Delta_{\text{D}} \big((-1 + e^{2 i d k_{z}}) \mu - (1 + e^{2 i d k_{z}})E_{5}(\boldsymbol{k})\big)}\Bigg),
\end{align}
\label{eigenvec1}
\end{subequations}
\begin{subequations}
\begin{align}
\Psi_{2,\boldsymbol{k}}&=\Bigg(\frac{
    A \sqrt{2(\Delta_{\text{D}}^2 + 
     2 \mu^2 + \Delta_{\text{D}}^2 \text{c}(2dk_{z}))} (i\text{s}k_{x} + 
     \text{s}k_{y})}{\Delta_{\text{D}} \big((-1 + e^{2idk_{z}}) \mu + (1 + 
        e^{2idk_{z}})E_{2}(\boldsymbol{k})\big)}, \,
-\frac{2 \Delta_{\text{s}}(k_{x},k_{y}) \mu + 2 \sqrt{\mu^2 + \Delta_{\text{D}}^2 \text{c}(dk_{z})^2} (\mu - E_{2}(\boldsymbol{k}))}{\Delta_{\text{D}} \big((-1 + e^{2idk_{z}}) \mu + (1 + e^{2idk_{z}}) E_{2}(\boldsymbol{k})\big)}, \nonumber\\
&\frac{2 A \mu (i \text{s}k_{x} + \text{s}k_{y})}{\Delta_{\text{D}}\big((-1 + e^{2idk_{z}}) \mu + (1 + e^{2idk_{z}}) E_{2}(\boldsymbol{k})\big)}, \nonumber\\
&\frac{((-1 + e^{2idk_{z}}) \Delta_{\text{D}}^2 \mu - 2A^2 \mu ((\text{s}k_{x})^{2} + (\text{s}k_{y})^{2}) - 
     2 \mu \Delta_{\text{S}}(k_{x},k_{y})^2 + 
     e^{idk_{z}} \Delta_{\text{D}}^2 \text{c}(dk_{z}) 2E_{2}(\boldsymbol{k}) + 
     \sqrt{(\mu^2 + \Delta_{\text{D}}^2 \text{c}(dk_{z})^2) \Delta_{\text{S}}(k_{x},k_{y})^2} (-2 \mu + 
        2E_{2}(\boldsymbol{k})))}{(\mu + E_{2}(\boldsymbol{k})) \Delta_{\text{D}} ((-1+e^{2idk_{z}})\mu+(1+e^{2idk_{z}})E_{2}(\boldsymbol{k}))}, \nonumber\\
&-\frac{(1 + e^{2idk_{z}}) A (-i \text{s}k_{x} - \text{s}k_{y})}{(-1 + e^{2idk_{z}}) \mu + (1 + e^{2idk_{z}})E_{2}(\boldsymbol{k})}, \, 1, \,0, \,
\frac{(1 + e^{2idk_{z}}) \Delta_{\text{s}}(k_{x},k_{y}) + 
  e^{2idk_{z}} \sqrt{2 (\Delta_{\text{D}}^2 + 2 \mu^2 + \Delta_{\text{D}}^2 \text{c}(2dk_{z}))}}{(-1 + e^{2idk_{z}}) \mu + (1 + e^{2idk_{z}}) E_{2}(\boldsymbol{k})}\Bigg), \nonumber\\
\Psi_{8,\boldsymbol{k}}&=\Bigg(\frac{
    A \sqrt{2(\Delta_{\text{D}}^2 + 
     2 \mu^2 + \Delta_{\text{D}}^2 \text{c}(2dk_{z}))} (i\text{s}k_{x} + 
     \text{s}k_{y})}{\Delta_{\text{D}} \big((-1 + e^{2idk_{z}}) \mu + (1 + 
        e^{2idk_{z}})E_{8}(\boldsymbol{k})\big)}, \,
-\frac{2 \Delta_{\text{s}}(k_{x},k_{y}) \mu + 2 \sqrt{\mu^2 + \Delta_{\text{D}}^2 \text{c}(dk_{z})^2} (\mu - E_{8}(\boldsymbol{k}))}{\Delta_{\text{D}} \big((-1 + e^{2idk_{z}}) \mu + (1 + e^{2idk_{z}}) E_{8}(\boldsymbol{k})\big)}, \nonumber\\
&\frac{2 A \mu (i \text{s}k_{x} + \text{s}k_{y})}{\Delta_{\text{D}}\big((-1 + e^{2idk_{z}}) \mu + (1 + e^{2idk_{z}}) E_{8}(\boldsymbol{k})\big)}, \nonumber\\
&\frac{((-1 + e^{2idk_{z}}) \Delta_{\text{D}}^2 \mu - 2A^2 \mu ((\text{s}k_{x})^{2} + (\text{s}k_{y})^{2}) - 
     2 \mu \Delta_{\text{S}}(k_{x},k_{y})^2 + 
     e^{idk_{z}} \Delta_{\text{D}}^2 \text{c}(dk_{z}) 2E_{8}(\boldsymbol{k}) + 
     \sqrt{(\mu^2 + \Delta_{\text{D}}^2 \text{c}(dk_{z})^2) \Delta_{\text{S}}(k_{x},k_{y})^2} (-2 \mu + 
        2E_{8}(\boldsymbol{k})))}{(\mu + E_{8}(\boldsymbol{k})) \Delta_{\text{D}} ((-1+e^{2idk_{z}})\mu+(1+e^{2idk_{z}})E_{8}(\boldsymbol{k}))}, \nonumber\\
&-\frac{(1 + e^{2idk_{z}}) A (-i \text{s}k_{x} - \text{s}k_{y})}{(-1 + e^{2idk_{z}}) \mu + (1 + e^{2idk_{z}})E_{8}(\boldsymbol{k})}, \, 1, \,0, \,
\frac{(1 + e^{2idk_{z}}) \Delta_{\text{s}}(k_{x},k_{y}) + 
  e^{2idk_{z}} \sqrt{2 (\Delta_{\text{D}}^2 + 2 \mu^2 + \Delta_{\text{D}}^2 \text{c}(2dk_{z}))}}{(-1 + e^{2idk_{z}}) \mu + (1 + e^{2idk_{z}}) E_{8}(\boldsymbol{k})}\Bigg),
\end{align}
\begin{align}
\Psi_{4,\boldsymbol{k}}&=\Bigg(-\frac{
    A \sqrt{2(\Delta_{\text{D}}^2 + 
     2 \mu^2 + \Delta_{\text{D}}^2 \text{c}(2dk_{z}))} (i\text{s}k_{x} + 
     \text{s}k_{y})}{\Delta_{\text{D}} \big((-1 + e^{2idk_{z}}) \mu + (1 + 
        e^{2idk_{z}})E_{4}(\boldsymbol{k})\big)}, \,
-\frac{2 \Delta_{\text{s}}(k_{x},k_{y}) \mu - 2 \sqrt{\mu^2 + \Delta_{\text{D}}^2 \text{c}(dk_{z})^2} (\mu - E_{4}(\boldsymbol{k}))}{\Delta_{\text{D}} \big((-1 + e^{2idk_{z}}) \mu + (1 + e^{2idk_{z}}) E_{4}(\boldsymbol{k})\big)}, \nonumber\\
&\frac{2 A \mu (i \text{s}k_{x} + \text{s}k_{y})}{\Delta_{\text{D}}\big((-1 + e^{2idk_{z}}) \mu + (1 + e^{2idk_{z}}) E_{4}(\boldsymbol{k})\big)}, \nonumber\\
&\frac{((-1 + e^{2idk_{z}}) \Delta_{\text{D}}^2 \mu - 2A^2 \mu ((\text{s}k_{x})^{2} + (\text{s}k_{y})^{2}) - 
     2 \mu \Delta_{\text{S}}(k_{x},k_{y})^2 + 
     e^{idk_{z}} \Delta_{\text{D}}^2 \text{c}(dk_{z}) 2E_{4}(\boldsymbol{k}) - 
     \sqrt{(\mu^2 + \Delta_{\text{D}}^2 \text{c}(dk_{z})^2) \Delta_{\text{S}}(k_{x},k_{y})^2} (-2 \mu + 
        2E_{4}(\boldsymbol{k})))}{(\mu + E_{4}(\boldsymbol{k})) \Delta_{\text{D}} ((-1+e^{2idk_{z}})\mu+(1+e^{2idk_{z}})E_{4}(\boldsymbol{k}))}, \nonumber\\
&-\frac{(1 + e^{2idk_{z}}) A (-i \text{s}k_{x} - \text{s}k_{y})}{(-1 + e^{2idk_{z}}) \mu + (1 + e^{2idk_{z}})E_{4}(\boldsymbol{k})}, \, 1, \,0, \,
\frac{(1 + e^{2idk_{z}}) \Delta_{\text{s}}(k_{x},k_{y}) - 
  e^{2idk_{z}} \sqrt{2 (\Delta_{\text{D}}^2 + 2 \mu^2 + \Delta_{\text{D}}^2 \text{c}(2dk_{z}))}}{(-1 + e^{2idk_{z}}) \mu + (1 + e^{2idk_{z}}) E_{4}(\boldsymbol{k})}\Bigg), \nonumber\\
\Psi_{6,\boldsymbol{k}}&=\Bigg(-\frac{
    A \sqrt{2(\Delta_{\text{D}}^2 + 
     2 \mu^2 + \Delta_{\text{D}}^2 \text{c}(2dk_{z}))} (i\text{s}k_{x} + 
     \text{s}k_{y})}{\Delta_{\text{D}} \big((-1 + e^{2idk_{z}}) \mu + (1 + 
        e^{2idk_{z}})E_{6}(\boldsymbol{k})\big)}, \,
-\frac{2 \Delta_{\text{s}}(k_{x},k_{y}) \mu - 2 \sqrt{\mu^2 + \Delta_{\text{D}}^2 \text{c}(dk_{z})^2} (\mu - E_{6}(\boldsymbol{k}))}{\Delta_{\text{D}} \big((-1 + e^{2idk_{z}}) \mu + (1 + e^{2idk_{z}}) E_{6}(\boldsymbol{k})\big)}, \nonumber\\
&\frac{2 A \mu (i \text{s}k_{x} + \text{s}k_{y})}{\Delta_{\text{D}}\big((-1 + e^{2idk_{z}}) \mu + (1 + e^{2idk_{z}}) E_{6}(\boldsymbol{k})\big)}, \nonumber\\
&\frac{((-1 + e^{2idk_{z}}) \Delta_{\text{D}}^2 \mu - 2A^2 \mu ((\text{s}k_{x})^{2} + (\text{s}k_{y})^{2}) - 
     2 \mu \Delta_{\text{S}}(k_{x},k_{y})^2 + 
     e^{idk_{z}} \Delta_{\text{D}}^2 \text{c}(dk_{z}) 2E_{6}(\boldsymbol{k}) - 
     \sqrt{(\mu^2 + \Delta_{\text{D}}^2 \text{c}(dk_{z})^2) \Delta_{\text{S}}(k_{x},k_{y})^2} (-2 \mu + 
        2E_{6}(\boldsymbol{k})))}{(\mu + E_{6}(\boldsymbol{k})) \Delta_{\text{D}} ((-1+e^{2idk_{z}})\mu+(1+e^{2idk_{z}})E_{6}(\boldsymbol{k}))}, \nonumber\\
&-\frac{(1 + e^{2idk_{z}}) A (-i \text{s}k_{x} - \text{s}k_{y})}{(-1 + e^{2idk_{z}}) \mu + (1 + e^{2idk_{z}})E_{6}(\boldsymbol{k})}, \, 1, \,0, \,
\frac{(1 + e^{2idk_{z}}) \Delta_{\text{s}}(k_{x},k_{y}) - 
  e^{2idk_{z}} \sqrt{2 (\Delta_{\text{D}}^2 + 2 \mu^2 + \Delta_{\text{D}}^2 \text{c}(2dk_{z}))}}{(-1 + e^{2idk_{z}}) \mu + (1 + e^{2idk_{z}}) E_{6}(\boldsymbol{k})}\Bigg).
\end{align}
\label{eigenvec2}
\end{subequations}
\normalsize
At least one component of each of the $\Psi_{n,\boldsymbol{k}}$ in Eqs.~(\ref{eigenvec1}) and (\ref{eigenvec2}) is nonzero at every $\boldsymbol{k}\in\text{BZ}$, thus $\forall\boldsymbol{k}\in\text{BZ}:\braket{\Psi_{n,\boldsymbol{k}}}{\Psi_{n,\boldsymbol{k}}}\neq0$. Indeed for every $\boldsymbol{k}\in\text{BZ}$ each of the vectors $\ket{\Psi_{1,\boldsymbol{k}}},\ldots,\ket{\Psi_{8,\boldsymbol{k}}}$ defined by Eqs.~(\ref{eigenvec1}) and (\ref{eigenvec2}) are energy eigenvectors of Bloch's form (they satisfy $\ket{\Psi_{n,\boldsymbol{k}+\boldsymbol{G}}}=\ket{\Psi_{n,\boldsymbol{k}}}$) and are smooth over BZ. These vectors can be normalized to obtain
\begin{align}
    \ket{\psi_{n,\boldsymbol{k}}}\equiv\frac{\ket{\Psi_{n,\boldsymbol{k}}}}{\braket{\Psi_{n,\boldsymbol{k}}}{\Psi_{n,\boldsymbol{k}}}}.
\end{align}
We denote the corresponding cell-periodic functions by $\ket{u_{n,\boldsymbol{k}}}$.
Thus we identify a smooth global Hamiltonian gauge $\mathfrak{u}$ pointwise by  $\mathfrak{u}_{\bm{k}}=(\ket{u_{n,\boldsymbol{k}}})_{n\in\{1,\ldots,8\}}$. 

We identify the components of the nonAbelian Berry connection induced by the gauge $\mathfrak{u}$ by
\begin{align}
    \xi^a_{nm}(\boldsymbol{k})\equiv\big(\xi^{\mathfrak{u}}(\boldsymbol{k})\big)^a_{nm}=i\braket{u_{n,\boldsymbol{k}}}{\partial_{a}u_{m,\boldsymbol{k}}}.
    \label{connection}
\end{align}
In principle, the components (\ref{connection}) can be obtained by explicitly taking $\boldsymbol{k}$-derivatives of Eqs.~(\ref{eigenvec1}) and (\ref{eigenvec2}) and employing some approximation for the $\bm{k}$-derivatives of the $\ket{\bar{u}_{(\alpha,\sigma),\bm{k}}}$ (for example, using the assumption of atomic-like WFs or of highly localized WFs as described above).
However, we will perform our calculations numerically. Some useful (numerically obtained) relations that arise in this gauge choice are
\begin{gather}
    \xi^a_{11}=-\xi^a_{22} ,\, \xi^a_{33}=-\xi^a_{44} ,\, \xi^a_{55}=-\xi^a_{66} ,\, \xi^a_{77}=-\xi^a_{88}, \nonumber\\
    \xi^a_{13}=-\xi^a_{42} ,\, \xi^a_{14}=\xi^a_{32} ,\, \xi^a_{57}=-\xi^a_{86} ,\, \xi^a_{58}=\xi^a_{76}, \nonumber\\
    \xi^a_{15}=-\xi^a_{62} ,\, \xi^a_{16}=\xi^a_{52} ,\, \xi^a_{17}=-\xi^a_{82} ,\, \xi^a_{18}=\xi^a_{72}, \nonumber\\
    \xi^a_{35}=-\xi^a_{64} ,\, \xi^a_{36}=\xi^a_{54} ,\, \xi^a_{37}=-\xi^a_{84} ,\, \xi^a_{38}=\xi^a_{74}.
    \label{connectionRelations}
\end{gather}
The relations (\ref{connectionRelations}) arise in both the approximation of atomic-like WFs ($\partial_{a}\ket{\bar{u}_{(\alpha,\sigma),\boldsymbol{k}}}\equiv 0$) and the approximation of highly localized WFs (\ref{uBarDeriv}) for $\boldsymbol{\delta}_{2}=\boldsymbol{\delta}_{3}=\frac{d}{2}\boldsymbol{z}$ and $\boldsymbol{\delta}_{0}=\boldsymbol{\delta}_{1}=-\frac{d}{2}\boldsymbol{z}$.
These relations are analogous to those presented in Eq.~(42) of Ref.~\cite{mahon2023reconciling} for the nonmagnetic case. Indeed each of the relations in Eq.~(\ref{connectionRelations}) have a natural analog in the nonmagnetic case. However, in the nonmagnetic case there are relations that have no analog here; in the Hamiltonian gauge employed in the nonmagnetic case \cite{mahon2023reconciling}, $\xi^a_{vv'}\propto\xi^a_{cc'}$ for various combinations of $v$, $v'$, $c$, and $c'$ (in particular, $\xi^a_{11}=\xi^a_{33}$, $\xi^a_{22}=\xi^a_{44}$ and $\xi^a_{12}=-\xi^a_{34}$, where bands 1 and 2 are filled and bands 3 and 4 are empty). 
In the AFM case we \textit{do not} have a relation between any $\xi^a_{vv}$ and $\xi^a_{cc}$, nor do we have a relation between $\xi^a_{12}$ and $\xi^a_{78}$ or between $\xi^a_{34}$ and $\xi^a_{56}$. As we later demonstrate, the lack of such relations will result in a qualitative distinction between the partitioning of the magnetoelectric response into atomic-like and itinerant contributions in the nonmagnetic and AFM cases.

In the AFM case there also emerges a property of certain $\xi^a_{nm}$ that is not present in the nonmagnetic case. 
It turns out that certain band components of the AFM $\xi^{a}$ are numerically nonnegligible only when $a=z$. In particular,  
\begin{align}
    \xi^a_{13}, \, \xi^a_{14}, \, \xi^a_{15}, \, \xi^a_{16}, \, \xi^a_{23}, \, \xi^a_{24}, \, \xi^a_{25}, \, \xi^a_{26}, \, \xi^a_{73}, \, \xi^a_{74}, \, \xi^a_{75}, \, \xi^a_{76}, \, \xi^a_{83}, \, \xi^a_{84}, \, \xi^a_{85}, \, \xi^a_{86}& \neq 0 \text{ only if } a=3.
    \label{connectionRelations1}
\end{align}

\subsection{Computation of $\alpha_{\text{CS}}$}
\label{Appendix:3DTightBinding2}
We can now directly compute the topological magnetoelectric coefficient in the gauge $\mathfrak{u}$, which is given by
    \begin{align}
	&\alpha^{\mathfrak{u}}_{\text{CS}}=-\frac{e^2}{2\hbar c}\epsilon^{abd}\int_{\text{BZ}}\frac{d^{3}k}{(2\pi)^3}\left(\sum_{vv'}\xi^a_{vv'}(\bm{k})\partial_b\xi^d_{v'v}(\bm{k})-\frac{2i}{3}\sum_{vv'v_1}\xi^a_{vv'}(\bm{k})\xi^b_{v'v_1}(\bm{k})\xi^d_{v_1v}(\bm{k})\right),
	\label{supplement:alphaCS}
    \end{align}
where $v,v',v_{1}\in\{1,2,3,4\}$ in this case. Again, the gauge $\mathfrak{u}$ is defined by Eqs.~(\ref{eigenvec1}) and (\ref{eigenvec2}). 
We evaluate this integral numerically and interpolate to convergence. 
Some results are listed in Table \ref{Table:alphaCSresults}.
To correspond with MBX materials we focus on the parameter regime $B/\Delta_{\text{S}}<0$, and arbitrarily choose $\Delta_{\text{S}}=190$ meV and $B=-25$ meV.
The values of $\alpha^{\mathfrak{u}}_{\text{CS}}$ listed in Table~\ref{Table:alphaCSresults} are independent of $A\in\mathbb{R}_{+}$, which is a parameter that is related to the single layer lattice constant that was introduced during regularization of the effective low-energy coupled Dirac cone model. 
Notably, when the values of the crystal parameters are varied such that the pairs of degenerate bands remain isolated from one another, the value to which $\alpha^{\mathfrak{u}}_{\text{CS}}$ evaluates is unchanged.
The findings listed in Table~\ref{Table:alphaCSresults} are consistent with Fig.~2 (a) of the main text.
Due to the large number of parameters in our model (Eq.~(\ref{H_AFM})) we do not extract the entire topological phase diagram. 
Via explicit calculation we have found that the value to which $\alpha^{\mathfrak{u}}_{\text{CS}}$ evaluates is unchanged regardless of whether we adopt either of the above described approximations of atomic-like WFs ($\partial_{a}\ket{\bar{u}_{(\alpha,\sigma),\boldsymbol{k}}}\equiv 0$) or of highly localized WFs (\ref{uBarDeriv}) with $\boldsymbol{\delta}_{2}=\boldsymbol{\delta}_{3}=\frac{d}{2}\boldsymbol{z}$ and $\boldsymbol{\delta}_{0}=\boldsymbol{\delta}_{1}=-\frac{d}{2}\boldsymbol{z}$.
\begin{table*}[h]
    \centering
    \begin{tabular}{||c c c c||}
            \hline
		\text{\quad} $J_{\text{S}}/\Delta_{\text{S}}$\text{\quad} & \text{\quad} $\Delta_{\text{D}}/\Delta_{\text{S}}$\text{\quad} & \text{\quad} $J_{\text{D}}/J_{\text{S}}$
            \text{\quad} & \text{\quad} $\alpha^{\mathfrak{u}}_{\text{CS}}$ $(\frac{e^2}{hc})$ \text{\quad} \\ [0.5ex] 
		\hline\hline
		0.8 & $\pm 0.25$ & \begin{tabular}{@{}c@{}}0.25 \\ 0.5\end{tabular} & \begin{tabular}{@{}c@{}}0 \\ 0\end{tabular} \\
		\hline
		0.8 & $\pm 1.25$ & \begin{tabular}{@{}c@{}}0.25 \\ 0.5\end{tabular} & \begin{tabular}{@{}c@{}} 1/2 \\ 1/2 \end{tabular} \\
		\hline
	\end{tabular} \text{\qquad} 
	\begin{tabular}{||c c c c||}
            \hline
		\text{\quad} $J_{\text{S}}/\Delta_{\text{S}}$\text{\quad} & \text{\quad} $\Delta_{\text{D}}/\Delta_{\text{S}}$\text{\quad} & \text{\quad} $J_{\text{D}}/J_{\text{S}}$
            \text{\quad} & \text{\quad} $\alpha^{\mathfrak{u}}_{\text{CS}}$ $(\frac{e^2}{hc})$ \text{\quad} \\ [0.5ex] 
		\hline\hline
		1.5 & $\pm 0.25$ & \begin{tabular}{@{}c@{}}0.25 \\ 0.5\end{tabular} & \begin{tabular}{@{}c@{}}1/2 \\ 0\end{tabular} \\
		\hline
		1.5 & $\pm 1.25$ & \begin{tabular}{@{}c@{}}0.25 \\ 0.5\end{tabular} & \begin{tabular}{@{}c@{}} -1/2 \\ 3/2 \end{tabular} \\
		\hline
	\end{tabular}\\
\caption{Values of $\alpha^{\mathfrak{u}}_{\text{CS}}$ obtained from numerical integration of Eq.~(\ref{supplement:alphaCS}) for $\Delta_{\text{S}}=190$ meV and $B=-25$ meV.}
\label{Table:alphaCSresults}
\end{table*}

\subsection{Partitioning of $\alpha_{\text{CS}}$ into atomic-like and itinerant contributions}
\label{Appendix:3DTightBinding3}
\subsubsection{Considerations for a general smooth global Hamiltonian gauge}
We begin with a discussion of the linear response of orbital magnetization in a band insulator that exhibits TRS. We then restrict focus to our model, Eq.~(\ref{H_AFM}). 
In general, the electronic orbital magnetization $\boldsymbol{M}$ can be partitioned into a sum of atomic-like $\overline{\boldsymbol{M}}$ and itinerant $\widetilde{\boldsymbol{M}}$ contributions,
\begin{align}
\boldsymbol{M}=\overline{\boldsymbol{M}}+\widetilde{\boldsymbol{M}}.
\end{align}
This decomposition applies both in the ground state, $\boldsymbol{M}^{(0)}=\overline{\boldsymbol{M}}^{(0)}+\widetilde{\boldsymbol{M}}^{(0)}$ \cite{Resta2005,Resta2006,Mahon2019}, and at linear response to an electric field, $\boldsymbol{M}^{(E)}=\overline{\boldsymbol{M}}^{(E)}+\widetilde{\boldsymbol{M}}^{(E)}$ \cite{Malashevich2010,Mahon2020}.
We previously \cite{mahon2023reconciling} expounded on this decomposition.
We now employ expressions that were previously derived \cite{Mahon2020} for the linear response of these contributions, and also use the results presented there that identify the terms in $\overline{\boldsymbol{M}}^{(E)}$ and $\widetilde{\boldsymbol{M}}^{(E)}$ that give rise to $\alpha_{\text{CS}}$ and $\alpha^{il}_{\text{G}}$.

We employ Eq.~(70) and Eqs.~(71) and (72) of Ref.~\cite{Mahon2020} for $\overline{\boldsymbol{M}}^{(E)}$ and $\widetilde{\boldsymbol{M}}^{(E)}$, respectively, which applies to any crystalline insulator whose electronic ground state is characterized by a vanishing (triple of) Chern number(s). We will initially work in a general smooth (and periodic) Hamiltonian gauge, such that $U_{n\alpha}(\boldsymbol{k})\equiv \delta_{n\alpha}$ in those expressions; the WFs with respect to which the electronic contribution to the electric polarization and orbital magnetization are identified are $\ket{W_{n,\boldsymbol{R}}}\propto\int_{\text{BZ}}d^{3}k e^{-i\boldsymbol{k}\cdot\boldsymbol{R}}\ket{\phi_{n,\boldsymbol{k}}}$. 
That is, we initially consider our gauge choice to be defined by unspecified Bloch energy eigenvectors $\ket{\phi_{n,\boldsymbol{k}}}$ that are smooth over BZ and only later will we take $\ket{\phi_{n,\boldsymbol{k}}}=\ket{\psi_{n,\boldsymbol{k}}}$ defined by Eqs.~(\ref{eigenvec1}) and (\ref{eigenvec2}). 
We denote the corresponding cell-periodic functions by $\ket{v_{n,\boldsymbol{k}}}$ and identify a smooth global Hamiltonian gauge $\mathfrak{v}$ pointwise by $\mathfrak{v}_{\bm{k}}=(\ket{v_{n,\boldsymbol{k}}})_{n\in\{1,\ldots,N\}}$ where $N$ is the number of energy bands in the model. 
We choose to work in a Hamiltonian gauge because we seek the existence of a smooth global gauge in which Eq.~(70) of Ref.~\cite{Mahon2020} vanishes, and in any Hamiltonian gauge the terms involving $\mathcal{W}^{a}$ vanish. Otherwise, those terms appear to be generically nonzero. Perhaps in special cases smooth global symmetric gauge choices exist (although in a $\mathbb{Z}_{2}$-odd phase a time-reversal symmetric smooth global frame is topologically forbidden \cite{Vanderbilt2011,Monaco2017}) for which these additional terms vanish.
Nevertheless, we focus on smooth global Hamiltonian gauge choices.
In such a gauge, Eq.~(70) of Ref.~\cite{Mahon2020} reduces to
\begin{align}
    \overline{M}_{\mathfrak{v}}^{i(E)}&=\frac{e^{2}}{4\hbar c}\epsilon^{iab}E^{l}\sum_{nm}f_{nm}\int_{\text{BZ}}\frac{d^{3}k}{(2\pi)^{3}}\Bigg(
    \frac{\partial_{b}(E_{n,\boldsymbol{k}}+E_{m,\boldsymbol{k}})}{E_{m,\boldsymbol{k}}-E_{n,\boldsymbol{k}}}(\xi^{\mathfrak{v}})^{a}_{nm}(\xi^{\mathfrak{v}})^{l}_{mn}+2\sum_{s}\frac{E_{s,\boldsymbol{k}}-E_{m,\boldsymbol{k}}}{E_{m,\boldsymbol{k}}-E_{n,\boldsymbol{k}}}\text{Re}\big[i(\xi^{\mathfrak{v}})^a_{ns}(\xi^{\mathfrak{v}})^b_{sm}(\xi^{\mathfrak{v}})^l_{mn}\big] \Bigg)\nonumber\\
    &=\frac{e^{2}}{4\hbar c}\epsilon^{iab}E^{l}\int_{\text{BZ}}\frac{d^{3}k}{(2\pi)^{3}}\Bigg(
    \sum_{nm}f_{nm}\frac{\partial_{b}(E_{n,\boldsymbol{k}}+E_{m,\boldsymbol{k}})}{E_{m,\boldsymbol{k}}-E_{n,\boldsymbol{k}}}(\xi^{\mathfrak{v}})^{a}_{nm}(\xi^{\mathfrak{v}})^{l}_{mn}+4\sum_{vv'c}\frac{E_{v',\boldsymbol{k}}-E_{v,\boldsymbol{k}}}{E_{c,\boldsymbol{k}}-E_{v,\boldsymbol{k}}}\text{Re}\big[i(\xi^{\mathfrak{v}})^a_{vv'}(\xi^{\mathfrak{v}})^b_{v'c}(\xi^{\mathfrak{v}})^l_{cv}\big] \nonumber\\
    &\qquad\qquad\qquad\qquad\qquad\qquad-4\sum_{vcc'}\frac{E_{c',\boldsymbol{k}}-E_{c,\boldsymbol{k}}}{E_{v,\boldsymbol{k}}-E_{c,\boldsymbol{k}}}\text{Re}\big[i(\xi^{\mathfrak{v}})^a_{cc'}(\xi^{\mathfrak{v}})^b_{c'v}(\xi^{\mathfrak{v}})^l_{vc}\big]\Bigg)\nonumber\\
    &+\frac{e^{2}}{2\hbar c}\epsilon^{iab}E^{l}\int_{\text{BZ}}\frac{d^{3}k}{(2\pi)^{3}}\text{Re}\Bigg[-2i\sum_{vv'c}(\xi^{\mathfrak{v}})^a_{cv'}(\xi^{\mathfrak{v}})^b_{v'v}(\xi^{\mathfrak{v}})^l_{vc}+\sum_{vc}(\partial_{a}(\xi^{\mathfrak{v}})^b_{cv})(\xi^{\mathfrak{v}})^l_{vc}\Bigg],
    \label{mBar}
\end{align}
where $\big(\xi^{\mathfrak{v}}\big)^a_{nm}(\bm{k})=i\braket{v_{n,\boldsymbol{k}}}{\partial_{a}v_{m,\boldsymbol{k}}}$ are the components of the nonAbelian Berry connection induced by the gauge $\mathfrak{v}$.
In what follows we will often not explicitly indicate the $\bm{k}$-dependence of quantities.
Now recall that the explicit form of $\alpha^{il}_{\text{G}}$ is given by 
\begin{align}
\alpha^{il}_{\text{G}}&=\frac{e^{2}}{4\hbar c}\epsilon^{iab}E^{l}\int_{\text{BZ}}\frac{d^{3}k}{(2\pi)^{3}}\Bigg(
    2\sum_{nm}f_{nm}\frac{\partial_{b}(E_{n,\boldsymbol{k}}+E_{m,\boldsymbol{k}})}{E_{m,\boldsymbol{k}}-E_{n,\boldsymbol{k}}}\xi^{a}_{nm}\xi^{l}_{mn}+4\sum_{vv'c}\frac{E_{v',\boldsymbol{k}}-E_{v,\boldsymbol{k}}}{E_{c,\boldsymbol{k}}-E_{v,\boldsymbol{k}}}\text{Re}\big[i\xi^a_{vv'}\xi^b_{v'c}\xi^l_{cv}\big] \nonumber\\
    &\qquad\qquad\qquad\qquad\qquad\qquad-4\sum_{vcc'}\frac{E_{c',\boldsymbol{k}}-E_{c,\boldsymbol{k}}}{E_{v,\boldsymbol{k}}-E_{c,\boldsymbol{k}}}\text{Re}\big[i\xi^a_{cc'}\xi^b_{c'v}\xi^l_{vc}\big]\Bigg),
    \label{alphaG}
\end{align}
which is gauge invariant, hence the lack of a gauge label.
In general, (generalized) TRS implies $\alpha^{il}_{\text{G}}=0$. In addition, note that in the second and third terms in the brackets of Eq.~(\ref{alphaG}), Berry connection matrix elements appear ($\xi^a_{vv'}$ and $\xi^a_{cc'}$, respectively) that can be independently varied by changing the crystal parameters of the Bloch Hamiltonian (while maintaining TRS). Thus, each of the terms in the brackets of Eq.~(\ref{alphaG}) must separately vanish if there is TRS.
This implies that if the insulator exhibits TRS then all but the last line in the final equality in Eq.~(\ref{mBar}) vanishes:
\begin{align}
    \overline{M}_{\mathfrak{v}}^{i(E)}&\overset{\mathrm{TRS}}{=\joinrel=}\frac{e^{2}}{2\hbar c}\epsilon^{iab}E^{l}\int_{\text{BZ}}\frac{d^{3}k}{(2\pi)^{3}}\text{Re}\Bigg[-2i\sum_{vv'c}(\xi^{\mathfrak{v}})^a_{cv'}(\xi^{\mathfrak{v}})^b_{v'v}(\xi^{\mathfrak{v}})^l_{vc}+\sum_{vc}(\partial_{a}(\xi^{\mathfrak{v}})^b_{cv})(\xi^{\mathfrak{v}})^l_{vc}\Bigg]\nonumber\\
    &=\frac{e^{2}}{2\hbar c}\epsilon^{iab}E^{l}\int_{\text{BZ}}\frac{d^{3}k}{(2\pi)^{3}}\text{Re}\Bigg[-i\sum_{vv'c}(\xi^{\mathfrak{v}})^a_{cv'}(\xi^{\mathfrak{v}})^b_{v'v}(\xi^{\mathfrak{v}})^l_{vc}+i\sum_{vcc'}(\xi^{\mathfrak{v}})^a_{cc'}(\xi^{\mathfrak{v}})^b_{c'v}(\xi^{\mathfrak{v}})^l_{vc}\Bigg]\nonumber\\
    &\equiv \bar{\alpha}_{\mathfrak{v}}^{li}E^{l}.
    \label{mBar1}
\end{align}
We can now implement any smooth global Hamiltonian gauge $\mathfrak{v}$ to evaluate Eq.~(\ref{mBar1}). 

Consider any two gauge choices $\mathfrak{u}$ and $\mathfrak{v}$ of this type, which are identified pointwise as $\mathfrak{u}_{\boldsymbol{k}}=(\ket{u_{n,\boldsymbol{k}}})_{n\in\{1,\ldots,N\}}$ and $\mathfrak{v}_{\boldsymbol{k}}=(\ket{v_{n,\boldsymbol{k}}})_{n\in\{1,\ldots,N\}}$, respectively, where $\ket{u_{n,\boldsymbol{k}}}$ and $\ket{v_{n,\boldsymbol{k}}}$ are the cell-periodic part of Bloch energy eigenvectors $\ket{\psi_{n,\boldsymbol{k}}}$ and $\ket{\phi_{n,\boldsymbol{k}}}$. Any sets $\{\ket{\phi_{n,\boldsymbol{k}}}\}_{n,\boldsymbol{k}}$ and $\{\ket{\psi_{n,\boldsymbol{k}}}\}_{n,\boldsymbol{k}}$ of orthogonal energy eigenvectors are related at each $\boldsymbol{k}\in\text{BZ}$ by a $\Gamma^{*}$-periodic unitary transformation $T(\boldsymbol{k})$ and so too are their cell-periodic parts,
\begin{align}
    \ket{v_{n,\boldsymbol{k}}}=\sum_{m=1}^{N}\ket{u_{m,\boldsymbol{k}}}T_{mn}(\boldsymbol{k}),
    \label{gaugeTransf}
\end{align}
where $T_{mn}(\boldsymbol{k})\equiv\big(T(\boldsymbol{k})\big)_{mn}$.
In the case of a band insulator with $M\leq N$ occupied bands, $T(\boldsymbol{k})$ is represented by a block-diagonal unitary matrix with blocks of size $M\times M$ and $(N-M)\times(N-M)$. Then
\begin{align}
    (\xi^{\mathfrak{v}})^{a}_{nm}(\boldsymbol{k})\equiv i\braket{v_{n,\boldsymbol{k}}}{\partial_{a}v_{m,\boldsymbol{k}}}=\sum_{r,s=1}^{N}T^{\dagger}_{nr}(\boldsymbol{k})\big((\xi^{\mathfrak{u}})^{a}_{rs}(\boldsymbol{k})+\mathcal{T}^{a}_{rs}(\boldsymbol{k})\big)T_{sm}(\boldsymbol{k}),
    \label{connectionGaugeTransf}
\end{align}
where $\mathcal{T}^{a}_{rs}(\boldsymbol{k})\equiv i\sum_{l=1}^{N}(\partial_{a} T_{rl}(\boldsymbol{k}))T^{\dagger}_{ls}(\boldsymbol{k})$ and $T^{\dagger}_{ls}(\boldsymbol{k})\equiv\big(T^{\dagger}(\boldsymbol{k})\big)_{ls}$. Note that since $T(\boldsymbol{k})$ is block-diagonal, so too is $\mathcal{T}^{a}(\boldsymbol{k})$. Thus, $(\xi^{\mathfrak{v}})^{a}_{vc}=\sum_{v'c'}T^{\dagger}_{v,v'}(\xi^{\mathfrak{u}})^{a}_{v'c'}T_{c'c}$. Plugging this into Eq.~(\ref{mBar1}) we find
\begin{align}
    \bar{\alpha}_{\mathfrak{v}}^{li}
    &=\frac{e^{2}}{2\hbar c}\epsilon^{iab}\int_{\text{BZ}}\frac{d^{3}k}{(2\pi)^{3}}\text{Re}\Bigg[-i\sum_{vv'c}(\xi^{\mathfrak{u}})^a_{cv'}\big((\xi^{\mathfrak{u}})^b_{v'v}+\mathcal{T}^{b}_{v'v}\big)(\xi^{\mathfrak{u}})^l_{vc}+i\sum_{vcc'}\big((\xi^{\mathfrak{u}})^a_{cc'}+\mathcal{T}^{a}_{cc'}\big)(\xi^{\mathfrak{u}})^b_{c'v}(\xi^{\mathfrak{u}})^l_{vc}\Bigg] \nonumber\\
    &=\bar{\alpha}_{\mathfrak{u}}^{li}+\frac{e^{2}}{2\hbar c}\epsilon^{iab}\int_{\text{BZ}}\frac{d^{3}k}{(2\pi)^{3}}\text{Re}\Bigg[-i\sum_{vv'c}(\xi^{\mathfrak{u}})^a_{cv'}\mathcal{T}^{b}_{v'v}(\xi^{\mathfrak{u}})^l_{vc}+i\sum_{vcc'}\mathcal{T}^{a}_{cc'}(\xi^{\mathfrak{u}})^b_{c'v}(\xi^{\mathfrak{u}})^l_{vc}\Bigg].
    \label{mBar2}
\end{align}
That is, calculations of $\bar{\alpha}^{li}_{\mathfrak{v}}$ can always be made using the components of the Berry connection $\xi^{\mathfrak{u}}$ induced by some reference gauge $\mathfrak{u}$ and the components of the $\mathcal{T}^{a}$ obtained from the unitary transformation $T$ that relates $\mathfrak{v}$ and $\mathfrak{u}$.


\subsubsection{Specific considerations for our model}
In past work \cite{mahon2023reconciling} we studied the nonmagnetic analog of the model employed here and considered whether there exists a gauge $\mathfrak{v}$ in which $\bar{\alpha}_{\mathfrak{v}}^{li}$ vanishes (independent of the choice of model parameters). That model has $N=4$ and we considered the initial electronic state of the material to be such that $M=2$ (i.e., $v\in\{1,2\}$ and $c\in\{3,4\}$). Consider general smooth global (and periodic) gauges $\mathfrak{u}$ and $\mathfrak{v}$ in the total Bloch bundle, the components of which are related pointwise as in Eq.~(\ref{gaugeTransf}).
We only consider gauges in the total Bloch bundle that decompose as a product of smooth gauges on the occupied and unoccupied subbundles thereof.
For gauges of this type, in the nonmagnetic model gauge transformations are represented by a $4\times 4$ block-diagonal unitary matrix consisting of two $2\times2$ unitary blocks. 
The question is whether or not one can choose the two $2\times 2$ blocks of $T$ in such a way that $\mathcal{T}_{vv'}^{a}$ and $\mathcal{T}_{cc'}^{a}$ are such that the integrand of Eq.~(\ref{mBar2}) vanishes.
(The integral as a whole may accidentally vanish in some gauge for a given set of material parameters due to fine tuning, but if the Hamiltonian parameters are slightly altered this fine tuning cancellation would be destroyed -- it would be more general if the integrand itself always vanished.)
Indeed in past work \cite{mahon2023reconciling} we explicitly found a gauge in which this occurred.

Let us now focus on the antiferromagnetic model.
We first employ our particular gauge choice $\mathfrak{u}$ defined by the Bloch energy eigenvectors in Eqs.~(\ref{eigenvec1}) and (\ref{eigenvec2}).
Considering the $i=l$ component of Eq.~(\ref{mBar1}) and using the relations (\ref{connectionRelations}) and (\ref{connectionRelations1}), we find
\begin{align}
    \bar{\alpha}^{ll}_{\mathfrak{u}}
    &=\frac{e^{2}}{2\hbar c}\epsilon^{lab}\int_{\text{BZ}}\frac{d^{3}k}{(2\pi)^{3}}\text{Re}\Bigg[\sum_{v=1}^{2}\sum_{c=7}^{8}\Bigg(-i\sum_{v'=1}^{2}(\xi^{\mathfrak{u}})^a_{cv'}(\xi^{\mathfrak{u}})^b_{v'v}+i\sum_{c'=7}^{8}(\xi^{\mathfrak{u}})^a_{cc'}(\xi^{\mathfrak{u}})^b_{c'v}\Bigg)(\xi^{\mathfrak{u}})^l_{vc}\nonumber\\
    &\qquad\qquad\qquad\qquad\qquad\quad+\sum_{v=3}^{4}\sum_{c=5}^{6}\Bigg(-i\sum_{v'=3}^{4}(\xi^{\mathfrak{u}})^a_{cv'}(\xi^{\mathfrak{u}})^b_{v'v}+i\sum_{c'=5}^{6}(\xi^{\mathfrak{u}})^a_{cc'}(\xi^{\mathfrak{u}})^b_{c'v}\Bigg)(\xi^{\mathfrak{u}})^l_{vc}\Bigg].
    \label{mBarDiagu}
\end{align}
For Eq.~(\ref{mBarDiagu}) to generally vanish over the parameter space of the model, either the terms in each of the $(\ldots)$ must cancel one another or the two $(\ldots)$ terms must cancel each other.
Since there is generally no relation between, for example, $(\xi^{\mathfrak{u}})^l_{17}$ and $(\xi^{\mathfrak{u}})^l_{35}$, the $(\ldots)$ terms cannot generally cancel each other, thus the second possibility is eliminated.
The first possibility does not occur either, since there is generally no relationship between $(\xi^{\mathfrak{u}})^a_{12}$ and $(\xi^{\mathfrak{u}})^a_{78}$ in the first $(\ldots)$ of Eq.~(\ref{mBarDiagu}) nor between $(\xi^{\mathfrak{u}})^a_{34}$ and $(\xi^{\mathfrak{u}})^a_{56}$ in the second $(\ldots)$.
(In the original gauge $\mathfrak{u}$ employed in the nonmagnetic case there was such a relationship between $(\xi^{\mathfrak{u}})^a_{v'v}$ and $(\xi^{\mathfrak{u}})^a_{cc'}$ for $v\neq v'$ and $c\neq c'$.) 

What about in some other smooth global Hamiltonian gauge $\mathfrak{v}$?
Due to the degeneracy in the model we employ, at each $\bm{k}\in\text{BZ}$ the transformation $T(\boldsymbol{k})$ is represented by a block-diagonal matrix consisting of four $2\times 2$ unitary blocks. Then
\begin{align}
    \bar{\alpha}^{ll}_{\mathfrak{v}}
    &=\frac{e^{2}}{2\hbar c}\epsilon^{lab}\int_{\text{BZ}}\frac{d^{3}k}{(2\pi)^{3}}\text{Re}\Bigg[\sum_{v=1}^{2}\sum_{c=7}^{8}\Bigg(-i\sum_{v'=1}^{2}(\xi^{\mathfrak{u}})^a_{cv'}\big((\xi^{\mathfrak{u}})^b_{v'v}+\mathcal{T}^{b}_{v'v}\big)+i\sum_{c'=7}^{8}\big((\xi^{\mathfrak{u}})^a_{cc'}+\mathcal{T}^{a}_{cc'}\big)(\xi^{\mathfrak{u}})^b_{c'v}\Bigg)(\xi^{\mathfrak{u}})^l_{vc}\nonumber\\
    &\qquad\qquad\qquad\qquad\qquad\quad+\sum_{v=3}^{4}\sum_{c=5}^{6}\Bigg(-i\sum_{v'=3}^{4}(\xi^{\mathfrak{u}})^a_{cv'}\big((\xi^{\mathfrak{u}})^b_{v'v}+\mathcal{T}^{b}_{v'v}\big)+i\sum_{c'=5}^{6}\big((\xi^{\mathfrak{u}})^a_{cc'}+\mathcal{T}^{a}_{cc'}\big)(\xi^{\mathfrak{u}})^b_{c'v}\Bigg)(\xi^{\mathfrak{u}})^l_{vc}\Bigg].
    \label{mBarDiag}
\end{align}
Again, the lack of a relation between, for example, $(\xi^{\mathfrak{u}})^l_{17}$ and $(\xi^{\mathfrak{u}})^l_{35}$, means that the two $(\ldots)$ terms cannot generally cancel one another. So for the integrand to vanish there must be cancellation within each $(\ldots)$. 
Using the relations $(\xi^{\mathfrak{u}})^a_{17}=-(\xi^{\mathfrak{u}})^a_{82}$ and $(\xi^{\mathfrak{u}})^a_{18}=(\xi^{\mathfrak{u}})^a_{72}$ given in (\ref{connectionRelations}), the first $(\ldots)$ term in Eq.~(\ref{mBarDiag}) can be rewritten as
\begin{align}
    &\epsilon^{lab}\int_{\text{BZ}}\frac{d^{3}k}{(2\pi)^{3}}\text{Re}\Bigg[\sum_{v=1}^{2}\sum_{c=7}^{8}\Bigg(-i\sum_{v'=1}^{2}(\xi^{\mathfrak{u}})^a_{cv'}\big((\xi^{\mathfrak{u}})^b_{v'v}+\mathcal{T}^{b}_{v'v}\big)+i\sum_{c'=7}^{8}\big((\xi^{\mathfrak{u}})^a_{cc'}+\mathcal{T}^{a}_{cc'}\big)(\xi^{\mathfrak{u}})^b_{c'v}\Bigg)(\xi^{\mathfrak{u}})^l_{vc}\Bigg]\nonumber\\
    &=\epsilon^{lab}\int_{\text{BZ}}\frac{d^{3}k}{(2\pi)^{3}}\text{Re}\Bigg[
    i(\xi^{\mathfrak{u}})^l_{17}\bigg(\Big\{(\xi^{\mathfrak{u}}+\mathcal{T})^{a}_{11}+(\xi^{\mathfrak{u}}+\mathcal{T})^{a}_{77}-(\xi^{\mathfrak{u}}+\mathcal{T})^{a}_{22}-(\xi^{\mathfrak{u}}+\mathcal{T})^{a}_{88}
    \Big\}(\xi^{\mathfrak{u}})^{b}_{71}\nonumber\\
    &\qquad\qquad\qquad\qquad\qquad\qquad\qquad+2(\xi^{\mathfrak{u}}+\mathcal{T})^{a}_{78}(\xi^{\mathfrak{u}})^{b}_{81}+2(\xi^{\mathfrak{u}}+\mathcal{T})^{a}_{12}(\xi^{\mathfrak{u}})^{b}_{18}\bigg)\nonumber\\
    &\quad\qquad\qquad\qquad\qquad+i(\xi^{\mathfrak{u}})^l_{18}\bigg(\Big\{(\xi^{\mathfrak{u}}+\mathcal{T})^{a}_{11}+(\xi^{\mathfrak{u}}+\mathcal{T})^{a}_{88}-(\xi^{\mathfrak{u}}+\mathcal{T})^{a}_{22}-(\xi^{\mathfrak{u}}+\mathcal{T})^{a}_{77}
    \Big\}(\xi^{\mathfrak{u}})^{b}_{81}\nonumber\\
    &\qquad\qquad\qquad\qquad\qquad\qquad\qquad+2(\xi^{\mathfrak{u}}+\mathcal{T})^{a}_{87}(\xi^{\mathfrak{u}})^{b}_{71}-2(\xi^{\mathfrak{u}}+\mathcal{T})^{a}_{21}(\xi^{\mathfrak{u}})^{b}_{17}\bigg)\Bigg],
    \label{mBarDiag1}
\end{align}
where we have adopted the shorthand $(\xi^{\mathfrak{u}}+\mathcal{T})^{a}_{nm}(\boldsymbol{k})\equiv(\xi^{\mathfrak{u}})^a_{nm}(\boldsymbol{k})+\mathcal{T}^{a}_{nm}(\boldsymbol{k})$.
The terms involving the $\{\ldots\}$ factors in Eq.~(\ref{mBarDiag1}) can simultaneously vanish only if there exists a $T$ such that $(\xi^{\mathfrak{u}}+\mathcal{T})^{a}_{11}=(\xi^{\mathfrak{u}}+\mathcal{T})^{a}_{22}$ and $(\xi^{\mathfrak{u}}+\mathcal{T})^{a}_{77}=(\xi^{\mathfrak{u}}+\mathcal{T})^{a}_{88}$; for $A,B,C,D\in\mathbb{C}$, the solution to the set of algebraic equations $A-B+C-D=0$ and $A-B-C+D=0$ is $A=B$ and $C=D$.
Recall from Eq.~(\ref{connectionRelations}) that $(\xi^{\mathfrak{u}})^a_{11}=-(\xi^{\mathfrak{u}})^a_{22}$ and $(\xi^{\mathfrak{u}})^a_{77}=-(\xi^{\mathfrak{u}})^a_{88}$.
For $n$ and $m$ in the same $2\times 2$ degenerate block of bands (for example, $n,m\in\{1,2\}$) we have $(T_{nm}(\bm{k}))_{n,m}\in U(2)$ and can always be written as $(T_{nm}(\bm{k}))_{n,m}=e^{-i\phi(\bm{k})}(S_{nm}(\bm{k}))_{n,m}$, where $(S_{nm}(\bm{k}))_{n,m}\in SU(2)$ is $\Gamma^{*}$-periodic and smooth in $\boldsymbol{k}$, and $\phi:\text{BZ}\rightarrow\mathbb{R}$ is a smooth map that satisfies $\phi(\bm{k}+\bm{G})=\phi(\bm{k})+2\pi N$ for some $N\in\mathbb{Z}$; we introduce independent $\phi$'s for each of the four degenerate blocks of bands. 
Then $\mathcal{T}^{a}_{nm}(\bm{k})=\delta_{nm}\partial_{a}\phi(\bm{k})+\mathcal{S}_{nm}^{a}(\bm{k})$, where $\mathcal{S}^{a}_{nm}\equiv i\sum_{l} (\partial_{a}S_{nl})S^{\dagger}_{lm}$ satisfies $\mathcal{S}^{a}_{11}=-\mathcal{S}^{a}_{22}$ since $S(\bm{k})\in SU(2)$ (i.e., we can write $S(\bm{k})=\begin{pmatrix}
a(\bm{k}) & b(\bm{k}) &\\
-b^*(\bm{k}) & a^*(\bm{k}) 
\end{pmatrix}$, where $a(\bm{k}),b(\bm{k})\in\mathbb{C}$ and $|a(\bm{k})|^{2}+|b(\bm{k})|^{2}=1$). 
This implies that the $\{\ldots\}$'s simultaneously vanish only if $(\xi^{\mathfrak{u}}+\mathcal{S})^{a}_{11}=(\xi^{\mathfrak{u}}+\mathcal{S})^{a}_{22}\equiv 0$ and $(\xi^{\mathfrak{u}}+\mathcal{S})^{a}_{77}=(\xi^{\mathfrak{u}}+\mathcal{S})^{a}_{88}\equiv 0$.


(In nonmagnetic case, these $\{\ldots\}$ terms could vanish for different reasons because of additional relations that are present in the original gauge $\mathfrak{u}$. 
In particular, in that case $(\xi^{\mathfrak{u}})^a_{13}$ (which roughly plays the role of the AFM $(\xi^{\mathfrak{u}})^a_{17}$ and $(\xi^{\mathfrak{u}})^a_{35}$) is real-valued, thus the analog of the term involving the first $\{\ldots\}$ is purely imaginary and thus gives no contribution. The analog of the second $\{\ldots\}$ term vanishes because $(\xi^{\mathfrak{u}})^a_{11}=-(\xi^{\mathfrak{u}})^a_{33}$ (the nonmagnetic $(\xi^{\mathfrak{u}})^a_{33}$ roughly plays the role of the AFM $(\xi^{\mathfrak{u}})^a_{77}$ and $(\xi^{\mathfrak{u}})^a_{55}$). Analogous relations do not arise in the AFM case.)


What about the other terms in Eq.~(\ref{mBarDiag1})?
For there to be general cancellation between the terms that do not involve $\{\ldots\}$ factors, we must assume that $T$ can additionally be chosen to relate $(\xi^{\mathfrak{u}}+\mathcal{T})^{a}_{12}$ and $(\xi^{\mathfrak{u}}+\mathcal{T})^{a}_{78}$; for example, $(\xi^{\mathfrak{u}}+\mathcal{T})^{a}_{12}\propto(\xi^{\mathfrak{u}}+\mathcal{T})^{a}_{78}$ or $(\xi^{\mathfrak{u}}+\mathcal{T})^{a}_{12}\propto(\xi^{\mathfrak{u}}+\mathcal{T})^{a}_{87}$.
However, since $\text{Re}[(\xi^{\mathfrak{u}})^a_{nm}]$, $\text{Im}[(\xi^{\mathfrak{u}})^a_{nm}]$ do not identically vanish for $n=1$ and $m=7$ or $8$, any such relations do not result in the vanishing of these other terms in Eq.~(\ref{mBarDiag1}).
Thus, the terms that do not involve $\{\ldots\}$ in Eq.~(\ref{mBarDiag1}) vanish only if $(\xi^{\mathfrak{u}}+\mathcal{T})^{a}_{12}=(\xi^{\mathfrak{u}}+\mathcal{T})^{a}_{78}\equiv 0$.

Using the relations $(\xi^{\mathfrak{u}})^a_{35}=-(\xi^{\mathfrak{u}})^a_{64}$ and $(\xi^{\mathfrak{u}})^a_{36}=(\xi^{\mathfrak{u}})^a_{54}$, and $(\xi^{\mathfrak{u}})^a_{33}=-(\xi^{\mathfrak{u}})^a_{44}$ and $(\xi^{\mathfrak{u}})^a_{55}=-(\xi^{\mathfrak{u}})^a_{66}$ given in Eq.~(\ref{connectionRelations}), analogous arguments apply to the the second bracketed term in Eq.~(\ref{mBarDiag}).
The sum total of these arguments is that there exists a smooth global Hamiltonian gauge $\mathfrak{v}$ in which $\bar{\alpha}^{ll}_{\mathfrak{v}}$ vanishes independent of material parameters only if there exists a $T$ relating smooth Hamiltonian gauges $\mathfrak{u}$ and $\mathfrak{v}$ that satisfies 
$(\xi^{\mathfrak{u}}+\mathcal{T})^{a}_{nm}=\delta_{nm}\partial_{a}\phi(\bm{k})$ (thus from Eq.~(\ref{connectionGaugeTransf}) we have $(\xi^{\mathfrak{v}})^{a}_{nm}=\delta_{nm}\partial_{a}\phi(\bm{k})$) for $n$ and $m$ in the same degenerate $2\times 2$ block of bands. 
And since $\mathfrak{v}$ is a Hamiltonian gauge (i.e., $T(\bm{k})$ and $\mathcal{T}^{a}(\bm{k})$ are block diagonal consisting of four $2\times 2$ blocks), Eq.~(\ref{connectionGaugeTransf}) implies that when $n$ and $m$ are not in the same degenerate $2\times 2$ block of bands, $(\xi^{\mathfrak{v}})^a_{nm}$ satisfies an analogous version of Eq.~(\ref{connectionRelations1}); that is, $(\xi^{\mathfrak{v}})^a_{13},\ldots,(\xi^{\mathfrak{v}})^a_{86}$ vanishes unless $a=z$.
If we \textit{assume} that such a $T$ (and therefore $\mathfrak{v}$ related to $\mathfrak{u}$ by that $T$) exists, then
\begin{align}
    \alpha^{\mathfrak{v}}_{\text{CS}}&=-\frac{e^2}{2\hbar c}\epsilon^{abd}\int_{\text{BZ}}\frac{d^{3}k}{(2\pi)^3}\left(\sum_{vv'}(\xi^{\mathfrak{v}})^a_{vv'}\partial_b(\xi^{\mathfrak{v}})^d_{v'v}-\frac{2i}{3}\sum_{vv'v_1}(\xi^{\mathfrak{v}})^a_{vv'}(\xi^{\mathfrak{v}})^b_{v'v_1}(\xi^{\mathfrak{v}})^d_{v_1v}\right)
    =0,
\end{align}
where we have used $\epsilon^{abd}\partial_b\partial_d \phi =0$ and $\epsilon^{abd}(\partial_b\phi)(\partial_d \phi) =0$.
This result is independent of material parameters.
Indeed this is a contradiction since we have explicitly shown above that this model supports regions of parameter space in which the electronic ground state (at half filling) is $\mathbb{Z}_{2}$-odd and $\alpha^{\mathfrak{v}}_{\text{CS}} \text{ mod } \frac{e^{2}}{hc}=\frac{e^{2}}{2hc}$. 
Thus, there does not exist a smooth global Hamiltonian gauge $\mathfrak{v}$ in which $\bar{\alpha}^{ll}_{\mathfrak{v}}$ vanishes independent of material parameters.
That is, in the bulk 3D AFM tight-binding model considered here, the magnetoelectric response cannot always be made entirely itinerant (with respect to a smooth global and periodic Hamiltonian gauge), in contrast to the nonmagnetic case \cite{mahon2023reconciling}.

\end{widetext}

\bibliography{MBT_TME}

\end{document}